\documentclass[aps,prd,showpacs,notitlepage,nofootinbib,floatfix,showkeys]{revtex4-2}

\usepackage{blindtext}
\usepackage{hyperref}
\usepackage{amsmath,amssymb}
\usepackage{float}
\usepackage{microtype}

\usepackage{graphicx}
\usepackage{bm}
\usepackage{latexsym}
\usepackage{epsfig}
\usepackage{psfrag}
\usepackage{color}
\usepackage[dvipsnames]{xcolor}
\usepackage{subfigure}
\usepackage{graphicx}

\usepackage{amssymb}
\usepackage{amsmath}
\usepackage{bm}
\usepackage{latexsym}
\usepackage{epsfig}
\usepackage{psfrag}
\usepackage[normalem]{ulem}
\usepackage{textcomp}
\usepackage{color}
\usepackage{pstricks}
\usepackage[utf8]{inputenc}

\def\nn{\nonumber} 

\def\f{\frac}
\def\l{\left}
\def\r{\right}
\def\d{{\rm d}}
\def\Mpl{M_{_{\mathrm{Pl}}}}
\def\mpcinv{{\rm Mpc}^{-1}}

\def\ps{\mathcal{P}_{_{\mathrm{S}}}}
\def\pc{\mathcal{P}_{_{\mathrm{C}}}}

\def\ph{\mathcal{P}_{h}}
\def\ns{n_{_{\mathrm{S}}}}

\def\cG{{\mathcal G}}
\def\cI{{\mathcal I}}

\def\cR{{\mathcal R}}

\def\cRG{{\mathcal R}^{^{_{\rm G}}}}
\def\cRk{{\mathcal R}_{\bm k}}
\def\cRGk{{\mathcal R}^{^{_{\rm G}}}_{\bm k}}

\def\ei{\eta_{\rm i}}

\def\ee{\eta_{\rm e}}

\def\ogw{\Omega_{_{\mathrm{GW}}}}
\def\fnl{f_{_{\rm NL}}}

\def\vka{{\bm k}_{1}}
\def\vkb{{\bm k}_{2}}
\def\vkc{{\bm k}_{3}}
\def\kf{k_{\rm peak}}

\def\vk{{\bm{k}}}
\def\vp{{\bm{p}}}
\def\vq{{\bm{q}}}

\def\d{{\mathrm{d}}}

\newcommand{\cf}{\textit{cf.~}}
\newcommand{\viz}{\textit{viz.~}}
\newcommand{\ie}{\textit{i.e.~}}

\allowdisplaybreaks

\begin{document}
\title{Accounting for scalar non-Gaussianity in secondary gravitational waves}
\author{H.~V.~Ragavendra}
\email{E-mail: ragavendra@physics.iitm.ac.in} 
\affiliation{Centre for Strings, Gravitation and Cosmology, 
Department of Physics, Indian Institute of Technology Madras, 
Chennai~600036, India}

\begin{abstract}
It is well known that enhancement in the primordial scalar perturbations 
over small scales generates detectable amplitudes of secondary gravitational 
waves (GWs), by sourcing the tensor perturbations at the second order. 
These stochastic gravitational waves are expected to carry the imprints of 
primordial non-Gaussianities.
The scalar bispectrum that is typically produced in models of inflation leading
to significant secondary GWs is non-trivial and highly scale-dependent.
In this work, we present a method to account for such general scale-dependent 
scalar bispectrum arising from inflationary models in the calculation of the 
spectral density of secondary GWs.
Using this method, we compute the contributions arising from the
scalar bispectrum to the amplitude of secondary GWs in two specific models of 
inflation driven by the canonical scalar field. We find that these non-Gaussian 
contributions can be highly model dependent and have to be consistently taken 
into account while estimating the total amplitude of the secondary GWs. 
Beyond the models considered, we emphasize that the method discussed is robust, 
free from assumptions about the shape of the bispectrum and generalizes earlier 
approaches adopted in the literature. 
We argue that this method of accounting for the scalar bispectrum shall be
helpful in future computations for exotic models generating larger amplitudes 
of scalar non-Gaussianities along with significant amount of secondary GWs.
\end{abstract}

\maketitle


\section{Introduction}

Models of inflation leading to enhanced scalar power over small scales are
examined in the context of production of primordial black holes (PBHs) and the 
associated secondary gravitational waves (GWs). In these models, modes of 
scalar perturbations that have amplitudes large enough to form PBHs, also 
enhance the tensor perturbations by sourcing them at the second order.
This leads to generation of secondary GWs of strengths detectable in the
present universe (see, for instance, 
Refs.~\cite{Ananda:2006af,Baumann:2007zm,Moore:2014lga,Garcia-Bellido:2017aan,
Bartolo:2018rku,Carr:2020gox} for discussions and constraints). 
Typical inflationary models considered in this context that 
are driven by a canonical scalar field, permit a brief epoch of ultra slow 
roll amidst an otherwise slow roll evolution of the inflaton field
(see, for instance, Refs.~\cite{Garcia-Bellido:2017mdw,Germani:2017bcs,
Dalianis:2018frf,Bhaumik:2019tvl,Ragavendra:2020sop}).
This epoch is known to enhance the amplitude of curvature perturbations and 
lead to large amplitudes of scalar power over small scales.
The production of PBHs is exponentially sensitive to the amplitude of scalar 
power and hence highly dependent on the behavior of the spectrum around the 
small range of wavenumbers close to the peak. However, the spectrum of 
secondary GWs is proportional to the square of the scalar power spectrum 
sourcing it. Therefore, it can better capture any feature that may be present 
in the scalar power spectrum over a wider range of wavenumbers.

Besides, there have been efforts to quantify the effect of the primordial
scalar non-Gaussianity on the predicted signals of secondary 
GWs~\cite{Cai:2018dig,Unal:2018yaa,Cai:2019amo,Ragavendra:2020vud,Atal:2021jyo,Adshead:2021hnm}. 
The general approach is to account 
for corrections in the power spectrum arising due to the scalar bispectrum 
through the non-Gaussianity parameter $\fnl$. There are usually well-motivated 
assumptions made about the shape of $\fnl$ being local in such calculations.
However, in realistic models of inflation, we find that, though $\fnl$ is local
close to the peak of the scalar power spectrum, it is highly scale dependent
over a wide range of wavenumbers.
Moreover, it has been shown that the consistency condition relating the power 
spectrum and the bispectrum in the squeezed limit is satisfied in canonical 
single field models considered in these scenarios~\cite{Ragavendra:2020sop,Zhang:2021vak}. 
Therefore, it is important to take into account the complete form of the 
bispectrum in calculating the correction to the power spectrum and examining 
the imprints of scalar non-Gaussianity on the secondary GWs.

In this work, we present a method to account for a general, scale-dependent
$\fnl$, in such a calculation, by reconsidering the definition of 
the parameter. This method does not assume any shape or template for $\fnl$ or 
the scalar bispectrum. Nevertheless it is consistent with the previous 
approaches when the assumptions are invoked, \ie it reduces to earlier methods
adopted if the $\fnl$ is assumed to be of a certain shape, say, a local form. 
This allows us to capture the complete behavior of the bispectrum along with
any non-trivial features that may be present therein and examine its imprint
on the spectrum of GWs generated. 
We illustrate this method of accounting for scalar bispectrum using two models 
as examples. One is a toy model of inflation constructed by adding an 
artificial dip to the otherwise smooth
potential permitting slow roll evolution of the 
field~\cite{Atal:2019cdz,Mishra:2019pzq}. 
The second is a model of inflation known as critical-Higgs inflation 
which is motivated by Higgs field driving inflation while containing an 
inflection point in the 
potential~\cite{Ezquiaga:2017fvi,Bezrukov:2017dyv,Drees:2019xpp}. 
Both these models serve as interesting examples for a typical scenario of 
inflation where the field undergoes an interim epoch of ultra slow roll during 
its evolution. We calculate the scalar bispectrum in these models and compute 
the corresponding correction to the power spectrum. We further compute the 
non-Gaussian contributions to the dimensionless spectral energy density of 
secondary GWs, \ie $\ogw$, generated in these models.

The structure of the paper is as follows. In the next section, we present the 
generalized definition of the non-Gaussianity parameter $\fnl$ to include a 
generic scale dependence. We also outline the steps involved in calculating 
$\fnl$ from the cubic order action of the scalar perturbations.
We then arrive at the expression for the correction to be added to the scalar 
power spectrum, \viz $\pc(k)$ in section~\ref{sec:pc}. 
In section~\ref{sec:ph-2}, we shall compute the non-Gaussian contributions to 
$\ogw$ arising due to $\fnl$. We shall point out that some of these 
contributions can be expressed in terms of $\pc(k)$.
We present the models for illustration in section~\ref{sec:models}, and compute 
the power and bi-spectra arising from them. We shall calculate the corrections 
to the power spectra using the respective bispectra and compare against the 
original spectra. We also obtain an analytical estimate of the correction
and compare it against the exact numerical result. We shall finally evaluate 
the $\ogw$ generated from these models due to both Gaussian and non-Gaussian 
contributions and compare the amplitudes in each case.
We conclude in section~\ref{sec:conc} with a brief summary and outlook.

Before proceeding, let us clarify the notations that shall be used in this
work. We work with natural units such that $\hbar = c = 1$ and set the reduced 
Planck mass to be $\Mpl=(8\,\pi\,G)^{-1/2}$.  We shall assume the background 
to be the spatially flat Friedmann-Lema\^itre-Robertson-Walker 
(FLRW) line element described by the scale factor $a$ and the Hubble parameter 
$H$. The Greek letter $\eta$ shall represent the conformal time coordinate.


\section{Scale dependent $\fnl(k_1,k_2,k_3)$}\label{sec:fnl-def}

In this section, we shall consider the conventional definition of the scalar
non-Gaussianity parameter $\fnl$ and generalize it to account for a generic
scale dependence. The parameter $\fnl$ is conventionally defined using the 
relation~\citep{Maldacena:2002vr, Martin:2011sn}
\begin{eqnarray}
\cR(\bm{x},\eta) &=& \cRG(\bm{x},\eta) - \f{3}{5}\fnl[\cRG(\bm{x},\eta)]^2\,,
\label{eq:fnl-std}
\end{eqnarray}
where $\cR(\bm{x},\eta)$ is the curvature perturbation and $\cRG(\bm{x},\eta)$ 
is the Gaussian part of $\cR(\bm{x},\eta)$.
Evidently, this definition assumes $\fnl$ to be local, \ie independent of
wavenumbers. Nevertheless, this is often taken as the definition to calculate 
the bispectrum even in cases with non-trivial scale dependence. Here, we shall 
generalize this definition to explicitly account for the scale dependence in 
the parameter. For this, we consider the above relation in Fourier space and 
redefine $\fnl$ as a function in Fourier space with wavenumbers as its 
arguments (for similar efforts in 
different contexts, see, Refs.~\cite{Schmidt:2010gw,Agullo:2021oqk}). We can 
write such a relation as
\begin{eqnarray}
\cRk(\eta) &=& \cRGk(\eta) - \f{3}{5}\int\f{\d^3\vka}{(2\,\pi)^{3/2}}
\cRG_{\bm{k_1}}(\eta)\cRG_{\bm{k}-\bm{k_1}}(\eta)\,
\fnl[\bm{k},(\vka-\bm{k}),-\vka]\,,
\label{eq:fnl-gen}
\end{eqnarray}
where $\cRk$ is the mode function corresponding to the curvature perturbation
$\cR$, and $\cRGk$ denotes the Gaussian part of~$\cRk$. 
We should mention that the $\fnl(k_1,k_2,k_3)$ defined depends only on 
the magnitude of the three wavevectors in the argument. We have written 
the arguments in the integrand above as vectors just to emphasize that by 
construction they form a triangular configuration in the space of wavenumbers 
(\ie sum of the three vectors vanishes identically), as is expected of the 
arguments of the bispectrum.
We can also obtain the counterpart of this parameter in real space by looking 
at the inverse Fourier transform of the above relation
\begin{eqnarray}
\cR(\bm{x},\eta) = \cRG(\bm{x},\eta) &-& \f{3}{5}\int\f{\d^3\bm{k}}{(2\pi)^3}
\int\d^3\vka \cRG_{\bm{k_1}}(\eta)\cRG_{\bm{k}-\bm{k_1}}(\eta)
\,\fnl[\bm{k},(\vka-\bm{k}),-\vka]\,{\rm e}^{i\bm{k}\cdot x}\,.
\end{eqnarray}
We should note that this equation reduces to the conventional definition of $\fnl$ 
given in Eq.~\eqref{eq:fnl-std}, if $\fnl(k_1,k_2,k_3)$ turns out to be scale 
independent in a given model. Hence our generalization is consistent with the
existing approach to quantify the scalar non-Gaussianity.

\subsection{Relation to the bispectrum}
We proceed to establish the relation between the $\fnl(k_1,k_2,k_3)$ given
above and the scalar bispectrum, denoted as $G(k_1,k_2,k_3)$.
Note that the scalar power spectrum $\ps(k)$ and the bispectrum 
$G(k_1,k_2,k_3)$ are defined as
\begin{subequations}
\begin{eqnarray}
\langle \hat{\cR}_{\bm{k_1}} \hat{\cR}_{\bm{k_2}} \rangle &=& 
\f{2\pi^2}{k^3}\ps(k_1)\,\delta^{(3)}(\vka + \vkb)\,, 
\label{eq:ps-def} \\
\langle \hat{\cR}_{\bm{k_1}} \hat{\cR}_{\bm{k_2}} \hat{\cR}_{\bm{k_3}} \rangle &=&
(2\pi)^{-3/2}\,G(k_1,k_2,k_3)\,\delta^{(3)}(\vka+\vkb+\vkc)\,,
\label{eq:G-def}
\end{eqnarray}
\end{subequations}
where $\hat\cRk$ is an operator obtained by quantizing the mode function $\cRk$.
To express $\fnl(k_1,k_2,k_3)$ in terms of $G(k_1,k_2,k_3)$ and $\ps(k)$, we 
compute the expectation value of the three point correlation of $\hat\cR_k$. 
Using the relation given in Eq.~\eqref{eq:fnl-gen}, we obtain that
\begin{eqnarray}
\langle \hat{\cR}_{\bm{k_1}} \hat{\cR}_{\bm{k_2}} \hat{\cR}_{\bm{k_3}} \rangle =
-\f{3}{5}\int & & \f{\d^3\vka'}{(2\,\pi)^{3/2}}\,
\langle \hat{\cal R}^{^{_{\rm G}}}_{\bm{k_1}} 
\hat{\cal R}^{^{_{\rm G}}}_{\bm{k_2}} 
\hat{\cal R}^{^{_{\rm G}}}_{\bm{k'_3}} 
\hat{\cal R}^{^{_{\rm G}}}_{\bm{k_3}-\bm{k'_3}} \rangle 
\nn \\
& &\times \,\fnl[\vkc,(\vkc'-\vkc),-\vkc'] + {\rm ~two~permutations}.
\end{eqnarray}
We should mention that the expectation values are evaluated in a specific 
initial state, which is assumed to be the Bunch-Davies vacuum. 
Also, note that the term in the right hand side of the above expression is 
the leading order term in the expansion assuming $\cRG$ is perturbative.
Using Wick's theorem, we can express the four point function in the above
integral in terms of the $\ps(k)$ and simplify it to obtain
\begin{eqnarray}
\langle \hat{\cR}_{\bm{k_1}} \hat{\cR}_{\bm{k_2}} \hat{\cR}_{\bm{k_3}} \rangle  &=& 
-\f{3}{5}\f{4\pi^4}{(2\pi)^{3/2}}\f{\ps(k_1)}{k_1^3}\f{\ps(k_2)}{k_2^3}
\,\delta^{(3)}(\vka+\vkb+\vkc) \nn \\
& & \times \bigg[ \fnl(\vkc,\vkb,\vka) + \fnl(\vkc,\vka,\vkb)\bigg]\,
+ {\rm ~two~permutations}.
\end{eqnarray}
We again emphasize that the arguments of $\fnl$ above are given as wavevectors
to remind that they satisfy the triangularity condition $\vka+\vkb+\vkc={\bf 0}$.
We then use the property of the bispectrum being symmetric in its arguments
[\ie $G(k_1,k_2,k_3) = G(k_1,k_3,k_2)$], to relate the $\fnl(k_1,k_2,k_3)$ 
constructed to the power and bi-spectra. We hence obtain the relation
\begin{eqnarray}
\fnl(k_1, k_2, k_3) &=& -\f{10}{3}\f{(k_1k_2k_3)^3}{16\pi^4}
G(k_1, k_2, k_3) 
\,[k_1^3\ps(k_2)\ps(k_3) + {\rm two~permutations}\,]^{-1}\,.
\label{eq:fnl-work}
\end{eqnarray}
This turns out to be the conventional relation used in the literature 
to compute $\fnl$ in terms of $\ps(k)$ and $G(k_1,k_2,k_3)$~\cite{Martin:2011sn,Hazra:2012yn,Sreenath:2014nca,Ragavendra:2020old}. 
Thus we infer that the $\fnl(k_1,k_2,k_3)$ defined in Eq.~\eqref{eq:fnl-gen} 
is compatible with the conventional relation. 
The difference in this derivation is that we have explicitly accounted for 
the scale dependence of the bispectrum in the non-Gaussianity 
parameter $\fnl(k_1,k_2,k_3)$.

\subsection{Calculation of $G(k_1,k_2,k_3)$ from the cubic order action}
Before we proceed to compute the correction to the power spectrum and the
non-Gaussian contributions to $\ogw$ due to $\fnl(k_1,k_2,k_3)$ discussed 
above, we shall briefly comment on the calculation of the scalar bispectrum 
$G(k_1,k_2,k_3)$ in a given inflationary model. 
The quantity $G(k_1,k_2,k_3)$, as defined in Eq.~\eqref{eq:G-def}, is 
evaluated in the perturbative vacuum, at the end of inflation. The bispectrum 
receives contributions from the third order action governing the curvature 
perturbation~\cite{Maldacena:2002vr,Seery:2005wm,Chen:2010xka,Collins:2011mz}. 
This action, in the case of a canonical scalar field driven inflation, has six 
terms bulk terms apart from boundary terms. Hence there arises six contributions 
to the bispectrum due to all the bulk terms. There is also a seventh 
contribution, arising due to a non-vanishing temporal boundary term, that is
typically absorbed using a field redefinition~\cite{Arroja:2011yj}.
The explicit forms of these contributions are as follows~\cite{Martin:2011sn,Hazra:2012yn,Ragavendra:2020old}: 
\begin{eqnarray}
G(\vka,\vkb,\vkc) 
&=& \sum_{C=1}^{7}\; G_{_{C}}(\vka,\vkb,\vkc)\nn\\
&=& \Mpl^2\; \sum_{C=1}^{6}\; 
\Biggl[f_{k_1}(\ee)\, f_{k_2}(\ee)\,f_{k_3}(\ee)\, 
\cG_{_{C}}(\vka,\vkb,\vkc)+{\mathrm{complex\;conjugate}}\Biggr]
+\, G_{7}(\vka,\vkb,\vkc)\,.\label{eq:sbs}
\end{eqnarray}
The terms denoted by $\cG_{_{C}}$ in this expression involve integrals 
arising from the bulk terms of the third order action. The seventh term $G_7$ 
is due to the non-vanishing boundary term mentioned above. 
The explicit forms of these terms are
\begin{subequations}\label{eq:cG}
\begin{eqnarray}
\cG_1(\vka,\vkb,\vkc)
&=& 2\,i\,\int_{\ei}^{\ee} \d\eta\; a^2\, 
\epsilon_{1}^2\, \l(f_{k_1}^{\ast}\,f_{k_2}'^{\ast}\,
f_{k_3}'^{\ast} + {\mathrm{two~permutations}}\r),\label{eq:cG1}\\
\cG_2(\vka,\vkb,\vkc)
&=&-2\,i\;\l(\vka\cdot \vkb + {\mathrm{two~permutations}}\r)\, 
\int_{\ei}^{\ee} \d\eta\; a^2\, 
\epsilon_{1}^2\, f_{k_1}^{\ast}\,f_{k_2}^{\ast}\,
f_{k_3}^{\ast},\label{eq:cG2}\\
\cG_3(\vka,\vkb,\vkc)
&=&-2\,i\,\int_{\ei}^{\ee} \d\eta\; a^2\,\epsilon_{1}^2\,
\l(\f{\vka\cdot\vkb}{k_2^2}\,
f_{k_1}^{\ast}\,f_{k_2}'^{\ast}\, f_{k_3}'^{\ast}
+ {\mathrm{five~permutations}}\r),\label{eq:cG3}\\
\cG_4(\vka,\vkb,\vkc)
&=& i\,\int_{\ei}^{\ee} \d\eta\; a^2\,\epsilon_{1}\,\epsilon_{2}'\, 
\l(f_{k_1}^{\ast}\,f_{k_2}^{\ast}\,f_{k_3}'^{\ast}
+{\mathrm{two~permutations}}\r),\label{eq:cG4}\\
\cG_5(\vka,\vkb,\vkc)
&=&\frac{i}{2}\,\int_{\ei}^{\ee} \d\eta\; 
a^2\, \epsilon_{1}^{3}\;\l(\f{\vka\cdot\vkb}{k_2^2}\,
f_{k_1}^{\ast}\,f_{k_2}'^{\ast}\, f_{k_3}'^{\ast} 
+ {\mathrm{five~permutations}}\r),\label{eq:cG5}\\
\cG_6(\vka,\vkb,\vkc) 
&=&\frac{i}{2}\,\int_{\ei}^{\ee}\d\eta\, a^2\, \epsilon_{1}^{3}\,
\l(\f{k_1^2\,\l(\vkb\cdot\vkc\r)}{k_2^2\,k_3^2}\, 
f_{k_1}^{\ast}\, f_{k_2}'^{\ast}\, f_{k_3}'^{\ast} 
+ {\mathrm{two~permutations}}\r),\qquad\label{eq:cG6}\\
G_{7}(\vka,\vkb,\vkc)
&=& -i\,\Mpl^2\,\l[f_{k_1}(\ee)\,f_{k_2}(\ee)\,f_{k_3}(\ee)\r]\nn\\
& &\times\, \biggl[a^2\epsilon_1\epsilon_{2}\,
f_{k_1}^{\ast}(\eta)\,f_{k_2}^{\ast}(\eta)\,f_{k_3}'^{\ast}(\eta) 
+\mathrm{two~permutations} \biggr]_{\eta_i}^{\ee}
+~\mathrm{complex~conjugate}\,,\label{eq:G7}
\end{eqnarray}
\end{subequations}
where in all these equations, $f_k(\eta)$ denotes the mode function
satisfying the Bunch-Davies initial condition and $f_k^\prime(\eta)$ 
is the derivative of $f_k(\eta)$ with respect to the conformal time $\eta$. 
The quantities $\epsilon_1$ and $\epsilon_2$ are the first and second slow roll 
parameters that capture the background evolution in a given model of inflation. 
Moreover, the time $\ei$ denotes the conformal time when the initial conditions 
are imposed on the modes, while $\ee$ denotes the conformal time close to the
end of inflation.

Using these expressions for the various contributions to the scalar bispectrum,
we can evaluate the complete bispectrum $G(k_1,k_2,k_3)$. Upon evaluating
the bispectrum, we can readily obtain the non-Gaussianity parameter 
$\fnl(k_1,k_2,k_3)$ through the relation given in Eq.~\eqref{eq:fnl-work}.

\section{Correction to the power spectrum}\label{sec:pc}

Having setup a method to account for a generic scale dependence in the 
non-Gaussianity parameter $\fnl(k_1,k_2,k_3)$, we shall now proceed to compute 
the non-Gaussian correction to the $\ps(k)$ arising due to the bispectrum.
To compute the correction, which we shall call as $\pc(k)$, we calculate the 
two point correlation of $\hat\cR_k$, using the relation given in 
Eq.~\eqref{eq:fnl-gen} as
\begin{eqnarray}
\langle \hat{\cR}_{\bm{k_1}} \hat{\cR}_{\bm{k_2}} \rangle &=& 
\langle \hat{\cal R}^{^{_{\rm G}}}_{\bm{k_1}} 
\hat{\cal R}^{^{_{\rm G}}}_{\bm{k_2}} \rangle
+ \f{9}{25}\int\f{\d^3\vka'}{(2\,\pi)^{3}}\int \d^3\vkb'
\langle \hat{\cal R}^{^{_{\rm G}}}_{\bm{k'_1}} 
\hat{\cal R}^{^{_{\rm G}}}_{\bm{k_1}-\bm{k'_1}} 
\hat{\cal R}^{^{_{\rm G}}}_{\bm{k'_2}} 
\hat{\cal R}^{^{_{\rm G}}}_{\bm{k_2}-\bm{k'_2}} \rangle
\nn \\
& & \times \,
\fnl\big(k_1 ,\vert \vka'-\vka \vert ,k_1' \big)\,
\fnl\big(k_2,\vert \vkb'-\vkb \vert , k_2' \big)\,.
\label{eq:Rk4}
\end{eqnarray}
On substituting the definition of power spectrum [\cf~Eq.~\eqref{eq:ps-def}] 
and expressing the four point correlation in terms of the two point correlations 
as before, the above equation leads to
\begin{eqnarray}
\ps^{^{\rm M}}(k) &=& \ps(k) + 
\f{9}{50\pi} k^3 \int \d^3\vka\,\f{\ps(k_1)}{k_1^3}
\f{\ps(\vert\bm{k}-\vka\vert)}{\vert \bm{k}-\vka \vert^3}
\,\fnl^2[k,\vert \vka-\bm{k} \vert, k_1]\,,
\end{eqnarray}
where $\ps(k)$ denotes the original power spectrum corresponding to the
Gaussian perturbations $\cR^{^{\rm G}}$ and $\ps^{^{\rm M}}(k)$ denotes the
spectrum with the non-Gaussian correction taken into account.
Therefore, we can identify the correction $\pc(k)$, that is to be added to 
the original spectrum $\ps(k)$, as
\begin{eqnarray}
\pc(k) &=& \f{9}{50\pi} k^3 \int \d^3\vka\,\f{\ps(k_1)}{k_1^3}
\f{\ps(\vert\bm{k}-\vka\vert)}{\vert \bm{k}-\vka \vert^3}
\,\fnl^2[{k},\vert \vka-\bm{k} \vert,k_1]\,.
\end{eqnarray}
We should note that there can be additional terms to this correction
which involve the irreducible part of the four point correlation, 
\viz the trispectrum of scalar 
perturbations~\cite{Cai:2019amo,Unal:2018yaa,Adshead:2021hnm}. 
Such terms shall receive contributions from higher order terms of the
action and hence will be at higher order in perturbations than the terms we 
are working with. We believe those terms are beyond the 
scope of this work. In our analysis we shall restrict ourselves to the terms 
of four point correlations reduced in terms of the power spectra.

To simplify the above expression for $\pc(k)$ we perform a suitable 
change of variables. Defining a variable $u = \vert \bm{k} -\vka \vert$, 
we get
\begin{eqnarray}
\pc(k) &=& \f{9}{25}\,k^2
\int_0^\infty \f{\d k_1}{k_1^2}\,\ps(k_1) 
\int_{\vert\bm{k}-\vka\vert}^{\vert\bm{k}+\vka\vert}
\f{\d u}{u^2}\ps(u)\,\fnl^2[k,u,k_1]\,.
\end{eqnarray}
Further introducing $x = k_1/k$ and $y = u/k$, we get,
\begin{eqnarray}
\pc(k) &=& \f{9}{25} 
\int_0^\infty \d x
\int_{\vert 1-x \vert}^{\vert 1 + x \vert}
\d y \f{\ps(kx)}{x^2}\,\f{\ps(ky)}{y^2}
\,\fnl^2[{k},{kx},{ky}]\,.
\label{eq:pc-fnl}
\end{eqnarray}
Again, we can notice that if $\fnl(k_1,k_2,k_3)$ turns out to be scale 
independent we recover the expression for $\pc(k)$ that is used in case 
of a local $\fnl$~\cite{Cai:2018dig,Unal:2018yaa,Cai:2019amo,Ragavendra:2020vud}. 
If we use the relation between $\fnl(k_1,k_2,k_3)$ and the power and 
bi-spectra [\cf Eq.~\eqref{eq:fnl-work}], we can write down $\pc(k)$ 
explicitly in terms of $G(k_1,k_2,k_3)$ and $\ps(k)$ as
\begin{eqnarray}
\pc(k) &=& \f{4\,k^{12}}{(2\pi)^8}
\int_{0}^{\infty} \d x\int^{1+x}_{\vert 1-x \vert} \d y\,
\f{x^4y^4}{\ps(kx)\,\ps(ky)}
\,G^2({k},{kx},{ky}) \,
\bigg[1 +x^3\f{\ps(k)}{\ps(kx)}
+y^3\f{\ps(k)}{\ps(ky)}\bigg]^{-2}\,.
\label{eq:pc-G}
\end{eqnarray}
We should mention here that, because of the well regulated nature of the 
integral involved, we shall use Eq.~\eqref{eq:pc-fnl} as the working definition 
for computing $\pc(k)$.

\section{Computation of $\ogw$ accounting for $\fnl$}\label{sec:ph-2}

Having obtained the correction to the power spectra, $\pc(k)$, we shall 
proceed to compute the non-Gaussian contributions to $\ogw$.
During the computation of $\ogw$ there may arise contributions from $\fnl$ 
other than from~$\pc(k)$. These are referred to as connected contributions in the 
literature~\cite{Cai:2018dig,Unal:2018yaa,Adshead:2021hnm}. We should note that 
there are arguments in the literature suggesting that these contributions vanish
identically when integrated over azimuthal angles involved in the corresponding 
integrals~\cite{Cai:2018dig,Cai:2019amo}. However, detailed calculations suggest 
that this may not be the case when accounted for exact dependence of the 
integrand over these angles appropriately~\cite{Adshead:2021hnm}. 
In this work, we shall compute all the terms involved while consistently 
accounting for a scale dependent $\fnl$ in them. We shall later compare the 
respective contributions against the contribution from the original power 
spectrum to the estimate of $\ogw$, when we consider specific models for
illustration.

To begin with, let us recall the calculation of the secondary tensor power 
spectrum in terms of the scalar power spectrum (for some of the earlier 
discussions, see Refs.~\cite{Ananda:2006af,Baumann:2007zm}; for some of the 
recent efforts, see, Refs.~\cite{Inomata:2016rbd,Garcia-Bellido:2017aan,
Espinosa:2018eve,Kohri:2018awv,Bartolo:2018rku,Pi:2020otn,Ragavendra:2020sop}).
The two point correlation of the secondary tensor perturbation 
$h_k^\lambda(\eta)$ is related to the scalar perturbation $\cRk$ as
\begin{eqnarray}
\langle \hat h^\lambda_{\vka}(\eta)\, \hat h^{\lambda'}_{\vkb}(\eta)\rangle 
&=& \f{16}{81}\f{1}{k_1k_2\eta^2}
\int \f{\d^3\vp}{(2\pi)^{3/2}} \int \f{\d^3\vp'}{(2\pi)^{3/2}}\, Q^\lambda(k_1,p)\,Q^{\lambda'}(k_2,p')\nn\\
& & \times \l[\cI_c\l(\f{p}{k_1},\f{\vert\vka-\vp\vert}{k_1}\r)\,
\mathrm{cos}\l(k_1\,\eta\r)
+\cI_s\l(\f{p}{k_1},\f{\vert\vka-\vp\vert}{k_1}\r)\,
\mathrm{sin}\l(k_1\,\eta\r)\r]\nn\\
& & \times \l[\cI_c\l(\f{p'}{k_2},\f{\vert\vkb-\vp'\vert}{k_2}\r)\,
\mathrm{cos}\l(k_2\,\eta\r)
+\cI_s\l(\f{p'}{k_2},\f{\vert\vkb-\vp'\vert}{k_2}\r)\,
\mathrm{sin}\l(k_2\,\eta\r)\r]\nn\\
& & \times\, \langle  \hat \cR_{\vp}\, \hat \cR_{\vka-\vp}\, 
\hat \cR_{\vp'}\, \hat \cR_{\vkb-\vp'}\rangle,\label{eq:h2-r4}
\end{eqnarray}
where the functions $\cI_{c,s}(u,v)$ arise due to the transfer function relating 
the Bardeen potential during the radiation dominated epoch to the primordial
curvature perturbation. The form of these functions are described as
\begin{subequations}\label{eq:cI}
\begin{eqnarray}
\cI_c(v,u) &=& -\f{27\,\pi}{4\,v^3\,u^3}\,
\Theta\l(v+u-\sqrt{3}\r)\, (v^2+u^2-3)^2,\\
\cI_s(v,u) &=& -\f{27}{4\,v^3\,u^3}\, (v^2+u^2-3)\,
\l[4\,v\,u+ (v^2+u^2-3)\;{\rm log}\,
\biggl\vert\f{3-(v-u)^2}{3-(v+u)^2}\biggr\vert\r],
\end{eqnarray}
\end{subequations}
where $\Theta(z)$ denotes the theta function.
The function $Q^\lambda(k,p)$ arises from the polarization
tensor associated with the tensor modes. It is given by
\begin{eqnarray}
Q^\lambda(k,p) 
= \begin{cases}
\l(\f{p}{k}\r)^2\, \f{\sin^2\theta}{\sqrt{2}}\,
\cos\,(2\phi), & \text{for}\,\lambda=+,\\
\l(\f{p}{k}\r)^2\, \f{\sin^2\theta}{\sqrt{2}}\,
\sin\,(2\phi), & \text{for}\,\lambda=\times,
\end{cases}
\end{eqnarray}
where $\theta$ is the polar angle and $\phi$ is azimuthal angle associated
with the wavevector $\vp$ with $\vk$ taken along the z-axis.
The four point correlation present in Eq.~\eqref{eq:h2-r4} is the term 
which shall give rise to Gaussian and non-Gaussian contributions.
On substituting the expression of $\cRk$ as given in Eq.~\eqref{eq:fnl-gen}
in each of the four mode functions of this term, we obtain a series
of terms with different powers of $\fnl$.
The terms that are independent of $\fnl$ are evidently the Gaussian
contributions. The terms with higher powers of $\fnl$ are the
non-Gaussian contributions. We shall first obtain the secondary tensor
power spectrum arising from the Gaussian contribution. 
Focusing on the terms independent of $\fnl$ and using Wick's theorem,
we can express the four point correlation in terms of the 
two point correlations. This leads to the following expression for 
secondary tensor power spectrum $\ph(k,\eta)$ in terms of the scalar 
power spectrum:
\begin{eqnarray}
\ph(k,\eta)
&=& 2\,\f{16}{81}\,\f{2\,\pi^2}{k^2\,\eta^2}
\int \f{\d^3 \bm k'}{(2\,\pi)^3}\, Q^\lambda(k,k')\, Q_\lambda(k,k')\,
\cI^2(k,k')
\,\f{k^3\,\ps(k')\,\ps(\vert\vk-\vk'\vert)}
{k'^3\,\vert\vk-\vk'\vert^3},
\end{eqnarray}
where the $\ps(k)$ denotes the Gaussian part of the scalar power 
spectrum.
We should note that $\ph(k,\eta)$ is averaged over 
the oscillations that occur on small time scales. 
Hence, the quantity $\cI(k,k')$ can be expressed as
\begin{equation}
\cI^2(k,k') = \l[ \cI_c^2\l( \f{k'}{k},\f{\vert \vk - \vk' \vert}{k} \r) 
+ \cI_s^2\l( \f{k'}{k}, \f{\vert \vk - \vk' \vert}{k} \r) \r].
\end{equation}
Notice that the contraction of $Q(k,k')$ over $\lambda$ implies summing over 
both polarizations. Utilizing this expression of $\ph(k,\eta)$, we may express 
the dimensionless spectral energy density of GWs associated 
with secondary tensor perturbations in the current universe,
\viz $\ogw(k)$, as~\cite{Espinosa:2018eve,Ragavendra:2020sop}
\begin{eqnarray}
h^2\,\ogw(k)
&=& \f{1.38\times10^{-5}}{24}\,(k^2\,\eta^2)\,\ph(k,\eta).
\label{eq:ogw-ph}
\end{eqnarray}
This corresponds to $\ogw$ arising from the Gaussian contribution. 
As we further compute the non-Gaussian contributions to $\ph(k)$, we shall 
utilize the above relation to compute the corresponding $\ogw$ as well.

As to the non-Gaussian contributions to $\ph(k)$, there shall be terms arising
from the four point correlation of Eq.~\eqref{eq:h2-r4} that contain $\fnl^2$ and
$\fnl^4$. Let us first consider the terms at the level of~$\fnl^2$. 
These terms can be understood as arising when we introduce $\fnl$, as defined 
in Eq.~\eqref{eq:fnl-gen}, in two of $\cRk$ terms of the four point correlation.
This gives rise to three types of contributions to $\ph(k)$, which we shall
refer to as $\ph^{(2-1)}(k),\ph^{(2-2)}(k)$ and $\ph^{(2-3)}(k)$. 
The exact expressions that describe these three contributions are given by
\begin{subequations}
\begin{eqnarray}
\ph^{(2-1)}(k) 
&=& 
2^5\,\f{16}{81}\,\f{9}{25}\,\f{(2\,\pi^2)^2}{(2\,\pi)^6}\,\f{1}{k^2\,\eta^2}\nn\\
& &\times\,\int \d^3 \bm{q}_1 \int \d^3 \bm{q}_2\, 
Q^{\lambda}(k,q_1)\,Q_{\lambda}(k,q_2)\,\cI(k,q_1)\,\cI(k,q_2)\nn\\
& &\times\, k^3\,\f{\ps(q_2)\,\ps(\vert{\vq_2+\vk}\vert)\,\ps(\vert\vq_1 - \vq_2\vert)}
{q_2^3\,\vert \vq_2 + \vk \vert^3\,\vert{\vq_1 - \vq_2}\vert^3}\nn\\
& &\times\, \fnl(q_1,q_2,\vert \vq_1-\vq_2 \vert)\, 
\fnl(\vert \vk - \vq_1 \vert, \vert \vq_2 - \vq_1 \vert, \vert \vk + \vq_2 \vert),
\label{eq:ph-2-1-exp} \\
\ph^{(2-2)}(k) 
&=& 
2^5\,\f{16}{81}\,\f{9}{25}\,\f{(2\,\pi^2)^2}{(2\,\pi)^6}\,\f{1}{k^2\,\eta^2}\,
\int \d^3 \bm{q}_1 \int \d^3 \bm{q}_2\,
Q^{\lambda}(k,q_1)\,Q_{\lambda}(k,q_1)\,\cI^2(k,q_1)\nn\\
& & \times\, k^3\,\f{\ps(\vert\vk-\vq_1\vert)\,\ps(q_2)\,
\ps(\vert\vq_1-\vq_2\vert)}
{q_2^3\,\vert\vk-\vq_1\vert^3\,\vert\vq_1-\vq_2\vert^3}
\fnl^2(q_1,q_2, \vert \vq_1-\vq_2 \vert)\nn\\
&=& 2^5\,\f{16}{81}\,\f{(2\,\pi^2)}{(2\,\pi)^3}\,\f{1}{k^2\,\eta^2}\nn\\ 
& &\times\,\int \d^3 \bm{q}_1\, Q^{\lambda}(k,q_1)\,Q_{\lambda}(k,q_1)\,
\cI^2(k,q_1)\,k^3\,\f{\pc(q_1)\,\ps(\vert\vk-\vq_1\vert)}
{q_1^3\,\vert\vk-\vq_1\vert^3},
\label{eq:ph-2-2-exp}\\
\ph^{(2-3)}(k) 
&=& 
2^5\,\f{16}{81}\,\f{9}{25}\,\f{(2\,\pi^2)^2}{(2\,\pi)^6}\,
\f{1}{k^2\,\eta^2}\nn\\
& &\times\,\int \d^3 \bm{q}_1 \int \d^3 \bm{q}_2\, 
Q^{\lambda}(k,q_1)\,Q_{\lambda}(k,q_2)\,
\cI(k,q_1) \cI(k,q_2)\nn\\
& & \times\, k^3\,\f{\ps(q_1)\,\ps(q_2)\,\ps(\vert \vk-\vq_1+\vq_2 \vert)}
{q_1^3\,q_2^3\,\vert \vk-\vq_1+\vq_2 \vert^3}\nn\\
& & \times\, \fnl(\vert \vk-\vq_1 \vert, q_2, \vert \vk-\vq_1+\vq_2 \vert)\,
\fnl(\vert \vk+\vq_2 \vert, q_1, \vert \vk_1-\vq_1+\vq_2 \vert).
\label{eq:ph-2-3-exp}
\end{eqnarray}\label{eq:ph-fnl2}
\end{subequations}
\\
We should note that the numerical factors preceding the integrals have been
retained in their specific forms to give an idea of the origin of these terms. 
For instance, the powers of $2$ arise from the various possible configurations 
of wavenumbers corresponding to a contribution. 
The fraction $9/25$ arises from the factor of $3/5$ present in the definition 
of $\fnl$, whereas the factors of $2\pi$ arise from the definition of 
power spectrum and the Fourier transformations.

Besides, we have used the definition of $\pc(k)$ in $\ph^{(2-2)}(k)$ to reduce the
first expression and obtain Eq.~\eqref{eq:ph-2-2-exp} [\cf Eq.~\eqref{eq:pc-fnl}].
Such a simplification is not possible with $\ph^{(2-1)}(k)$ or $\ph^{(2-3)}(k)$.
It is useful to note that one can construct Feynman diagrams to represent these 
integrals (see, for instance, 
Refs.~\cite{Unal:2018yaa,Atal:2021jyo,Adshead:2021hnm}).
If we identify the diagrams with the above integrals, we find that 
$\ph^{(2-1)}(k)$ arises from what is called the C-type diagram, whereas
$\ph^{(2-3)}(k)$ arises from the Z-type diagram. The term $\ph^{(2-2)}(k)$ 
arises from what is known as the hybrid diagram (see, 
App.~\ref{app:loops} for a discussion on these diagrams). 
The difference between the integrals presented here and the corresponding
ones in the literature is the dependence of $\fnl$ over wavenumbers.
As mentioned earlier, it has been argued that the terms $\ph^{(2-1)}(k)$ 
and $\ph^{(2-3)}(k)$ shall vanish when integrated over the azimuthal angles 
and it is only the $\ph^{(2-2)}(k)$ term that survives~\cite{Cai:2018dig,
Cai:2019amo}.
However, it was later shown that $\ph^{(2-1)}(k)$ and $\ph^{(2-3)}(k)$ do 
not necessarily vanish when the angular dependences are appropriately 
accounted for~\cite{Adshead:2021hnm}.

Next, we shall consider contributions to $\ph(k)$ at the level of $\fnl^4$.
These terms can be understood as arising from $\fnl$ in all four of $\cRk$
in the four point function in Eq.~\eqref{eq:h2-r4}. 
In such a case, we obtain three contributions, which we shall call 
$\ph^{(4-1)}(k), \ph^{(4-2)}(k)$ and $\ph^{(4-3)}(k)$. 
The expressions describing these contributions are given by
\begin{subequations}
\begin{eqnarray}
\ph^{(4-1)}(k) 
&=& 
2^7\,\f{16}{81}\,\l(\f{9}{25}\r)^2\,\f{(2\,\pi^2)^3}{(2\,\pi)^9}\,
\f{1}{k^2\,\eta^2}
\int\d^3 \bm{q}_1 \int\d^3 \bm{q}_2\, Q^{\lambda}(k,q_1)\,Q_{\lambda}(k,q_2)\nn\\
& & \times\, \cI(k,q_1)\, \cI(k,q_2)\, k^3\, \int \d^3 \bm{q}'_2\, 
\f{\ps(\vert \vk-\vq_1+\vq_2-\vq'_2 \vert)}
{\vert \vk-\vq_1+\vq_2-\vq'_2 \vert^3}\, \nn\\
& & \times
\f{\ps(q'_2)\,\ps(\vert \vq_2-\vq'_2 \vert)\ps(\vert \vq_1+\vq'_2 \vert)}
{{q'_2}^3\, \vert \vq_2-\vq'_2 \vert^3\vert\, \vq_1+\vq'_2 \vert^3}\nn\\
& &\times\, \fnl(q_1, \vert -\vq'_2 \vert, \vert \vq_1+\vq'_2 \vert)\,
\fnl(\vert \vk-\vq_1 \vert, \vert \vk-\vq_1+\vq_2-\vq'_2 \vert, \vert \vq'_2-\vq_2 \vert)\nn\\
& &\times\, 
\fnl(q_2, q'_2, \vert \vq_2-\vq'_2 \vert)\,
\fnl(\vert \vk+\vq_2 \vert, \vert \vq_1-\vk-\vq_2+\vq'_2 \vert, \vert -\vq_1-\vq'_2 \vert)
\label{eq:ph-4-1} \\
\ph^{(4-2)}(k) 
&=& 2^7\,\f{16}{81}\,\f{(2\,\pi^2)}{(2\,\pi)^3}\,\f{1}{k^2\,\eta^2}
\,\int \d^3 \bm{q}_1\, Q^{\lambda}(k,q_1)\,Q_{\lambda}(k,q_1)\,
\cI^2(k,q_1)\,k^3\,\f{\pc(q_1)\,\pc(\vert \vk-\vq_1 \vert)}
{q_1^3\,\vert \vk-\vq_1 \vert^3},
\label{eq:ph-4-2}\\
\ph^{(4-3)}(k) 
&=& 2^7\,\f{16}{81}\,\l(\f{9}{25}\r)^2\,\f{(2\,\pi^2)^3}{(2\,\pi)^9}\,
\f{1}{k^2\,\eta^2}\, 
\int \d^3 \vq_1 \int \d^3 \vq'_1\, Q^{\lambda}(k,q_1)\, \cI(k,q_1)\nn\\
& & \times\, k^3\,\f{\ps(q'_1)\,\ps(\vert \vq_1-\vq'_1 \vert)\,
\ps(\vert \vk-\vq_1+\vq'_1)}
{{q'_1}^3\,\vert \vq_1-\vq'_1 \vert^3\,\vert\vk-\vq_1+\vq'_1 \vert^3}\nn\\
& & \times \fnl(q_1, q'_1, \vert \vq_1-\vq'_1 \vert)\,
\fnl(\vert \vk-\vq_1 \vert, \vert -\vq'_1 \vert, \vert \vk-\vq_1+\vq'_1 \vert) \nn\\
& & \times \int \d^3 \vq_2\, Q_{\lambda}(k,q_2)\,\cI(k,q_2)\,
\f{\ps(\vert \vq'_1-\vq_1-\vq_2 \vert)}{\vert \vq'_1-\vq_1-\vq_2 \vert^3}\nn\\
& & \times\, \fnl(q_2, \vert \vq_1+\vq_2-\vq'_1 \vert, \vert \vq'_1-\vq_1 \vert)\,
\fnl(\vert -\vk-\vq_2 \vert, \vert \vq'_1-\vq_1-\vq_2 \vert, 
\vert \vq_1-\vk-\vq'_1 \vert).
\label{eq:ph-4-3}
\end{eqnarray}
\end{subequations}
Once again, we have retained the numerical factors to understand
the origin of these factors.
Also, we have used the definition of $\pc(k)$ to reduce the expression
of $\ph^{(4-2)}(k)$ in terms of $\pc(k)$ [\cf Eq.~\eqref{eq:pc-fnl}]. 
This is known as the reducible contribution. 
The other two terms, \viz $\ph^{(4-1)}(k)$ and $\ph^{(4-3)}(k)$, cannot be 
rewritten in terms of $\pc(k)$ and they correspond to so-called
non-planar and planar Feynman diagrams, respectively 
(\cf App.~\ref{app:loops}; for a discussion in this context, also
see Ref.~\cite{Adshead:2021hnm}).

We can now utilize Eq.~\eqref{eq:ogw-ph} to compute the $\ogw$ arising 
from each of these non-Gaussian contributions as well as the Gaussian
contribution, and compare them against one another. 
However, we should note here that, the terms denoted as $\ph^{(2-i)}(k)$, 
containing $\fnl^2$, involve computation of six dimensional integrals 
and the terms denoted as $\ph^{(4-i)}(k)$, containing $\fnl^4$, involve 
performing nine dimensional integrals. 
Evidently, when we need to compute such integrals numerically, 
simpler methods such as the Boole's rule on a grid based sampling can be 
disadvantageous.
Also, in such conventional methods, one will require enormous number of 
sampling points to achieve reasonable level of convergence of integrals 
in higher dimensions.
Hence, one should resort to Monte-Carlo method of integration which
circumvents the issue of dimensionality with reasonable number of 
points~\cite{Press:2007nrf}.
Moreover, at each point of these integrals we require the power spectra and
$\fnl$ to be evaluated, with their respective dependences on wavenumbers. 
Therefore, arriving at numerical estimates of these non-Gaussian contributions 
for a case of inflation driven by non-trivial potentials is a computationally 
intensive exercise. 
There has been an earlier attempt in the literature to compute these 
contributions~\cite{Adshead:2021hnm}. 
But, we should point out that, in such efforts, the computations involved 
using analytical templates for the power spectra, such as the Dirac delta 
function or a lognormal function. 
Also, the $\fnl$ was assumed to be of local form with a given amplitude and 
without any scale dependence.
Hence, the computation of integrals in such cases is relatively easier.
However, in this work we compute both the power spectra 
and the $\fnl$ numerically from the action governing the perturbations for a 
given model of interest.
Therefore, the computation becomes significantly more intensive and hence 
takes considerably more time and processing power.
Due to this complexity in computation and constraints in implementation, in 
this work, we shall restrict ourselves to calculating the non-Gaussian 
contributions up to the level of $\fnl^2$, \ie terms denoted as $\ph^{(2-i)}$.


\section{Models for illustration}\label{sec:models}

In this section, we shall illustrate the calculation of the correction to the 
scalar power spectrum and the non-Gaussian contributions to $\ogw$ due to a
generic $\fnl(k_1,k_2,k_3)$ using two models of inflation. These models serve 
as good examples of a typical scenario of inflation leading to generation of 
secondary GWs of significant strengths.
These models permit a brief epoch of ultra slow roll leading to enhancement of 
scalar power over small scales. These scalar perturbations source the secondary
tensor perturbations and hence amplify the strength of secondary GWs over 
frequencies corresponding to those scales. 

The first model we shall consider is inflation driven by a potential which 
has a dip introduced to it by hand. Such scenarios where a bump or a dip
introduced in a rather smooth potential have been discussed in the literature 
in the context of PBH formation~\cite{Atal:2019cdz,Mishra:2019pzq}. 
Though it may not be well motivated or immediately realized from a high energy 
theory, it is a toy model that helps achieve a brief epoch of ultra slow roll 
during inflation and hence enhance the scalar power. Here we shall work with 
such a toy model consisting of a dip added to the well known potential of 
Starobinsky model. The form of this potential shall be 
\begin{eqnarray}
V(\phi) &=& V_0\,
\left[1 - \exp\left(-\sqrt{\f{2}{3}}\f{\phi}{\Mpl}\right)\right]^2\,
\left\{ 1 - \lambda\exp\left[-\f{1}{2}
\left(\f{\phi - \phi_0}{\Delta\phi}\right)^2\right] \right\}\,,
\label{eq:phi-dip}
\end{eqnarray}
where clearly the first part is the potential corresponding to 
Starobinsky model, while the second part in curly braces is the Gaussian
shaped dip located at $\phi_0$ having a coupling strength $\lambda$ 
and a width $\Delta\phi$ .
The values of the parameters involved are set to be 
$V_0=2.25\times10^{-10}\,\Mpl^4$, $\lambda = 2.58\times10^{-3}$,
$\phi_0 = 4.25\,\Mpl$ and $\Delta\phi= 2.8\times10^{-2}\,\Mpl$.
With the initial value of $\phi_{\rm i}=5.6\,\Mpl$, we achieve about 
$81$~e-folds of inflation with the epoch of ultra slow roll occurring 
at around $50$~e-folds from the beginning, as the field crosses and evolves 
beyond $\phi_0$. We shall refer to this model as SMD, standing for Starobinsky
model with a dip.

Another model we shall consider to illustrate our arguments is a model
known as critical-Higgs 
inflation~\cite{Ezquiaga:2017fvi,Bezrukov:2017dyv,Drees:2019xpp}. 
This model arises when the Higgs field is non-minimally coupled to gravity.
The effective potential in this scenario contains a point of inflection 
which leads to an epoch of ultra slow roll thereby enhancing the scalar 
power. The potential describing this model can be written as
\begin{eqnarray}
V(\phi) &=& V_0\,\f{\l[1+a\,\l(\mathrm{ln}\,z\r)^2\r]\,z^4}
{\l[1+c\,\l(1+b\,\mathrm{ln}\,z\r)\,z^2\r]^2}\,,
\label{eq:phi-Higgs}
\end{eqnarray}
where the quantity $z = \phi/\mu$. We shall choose the values of the 
parameters to be $\mu=1\,\Mpl$ and $V_0 = 1.5\times 10^{-8}\,\Mpl^4$.
The other parameters involved are related to each other through the 
inflection point $z_{\rm c}$ as follows:
\begin{subequations}
\begin{eqnarray}
a &=& \f{4}{1 + c\,z_{\rm c}^2 + 2\,\log(z_{\rm c}) - 
4\,\log^2(z_{\rm c})}\,,\\
b &=& 2\,\f{1 + c\,z_{\rm c}^2 + 4\,\log(z_{\rm c}) + 
2\,c\,z_{\rm c}^2\,\log{z_{\rm c}}}
{c\,z_{\rm c}^2\,[1 + c\,z_{\rm c}^2 + 2\,\log(z_{\rm c}) - 
4\,\log^2(z_{\rm c})]}\,.
\end{eqnarray}
\end{subequations}
We have set $\{c, z_{\rm c}\}=\{2.850, 0.784\}$ and arrived at $\{a, b\}$ 
using these values. For these values of the model parameters, and with an 
initial value of field $\phi_{\rm i} = 6.0\,\Mpl$, we achieve about 
$66$~e-folds of inflation.
The epoch of ultra slow roll occurs at around $31$ e-folds from the beginning of 
evolution as the field crosses the inflection point at $0.784\,\Mpl$.
We shall denote this model as CHI.

\begin{figure}[!t]
\includegraphics[width=7.5cm, height=5cm]{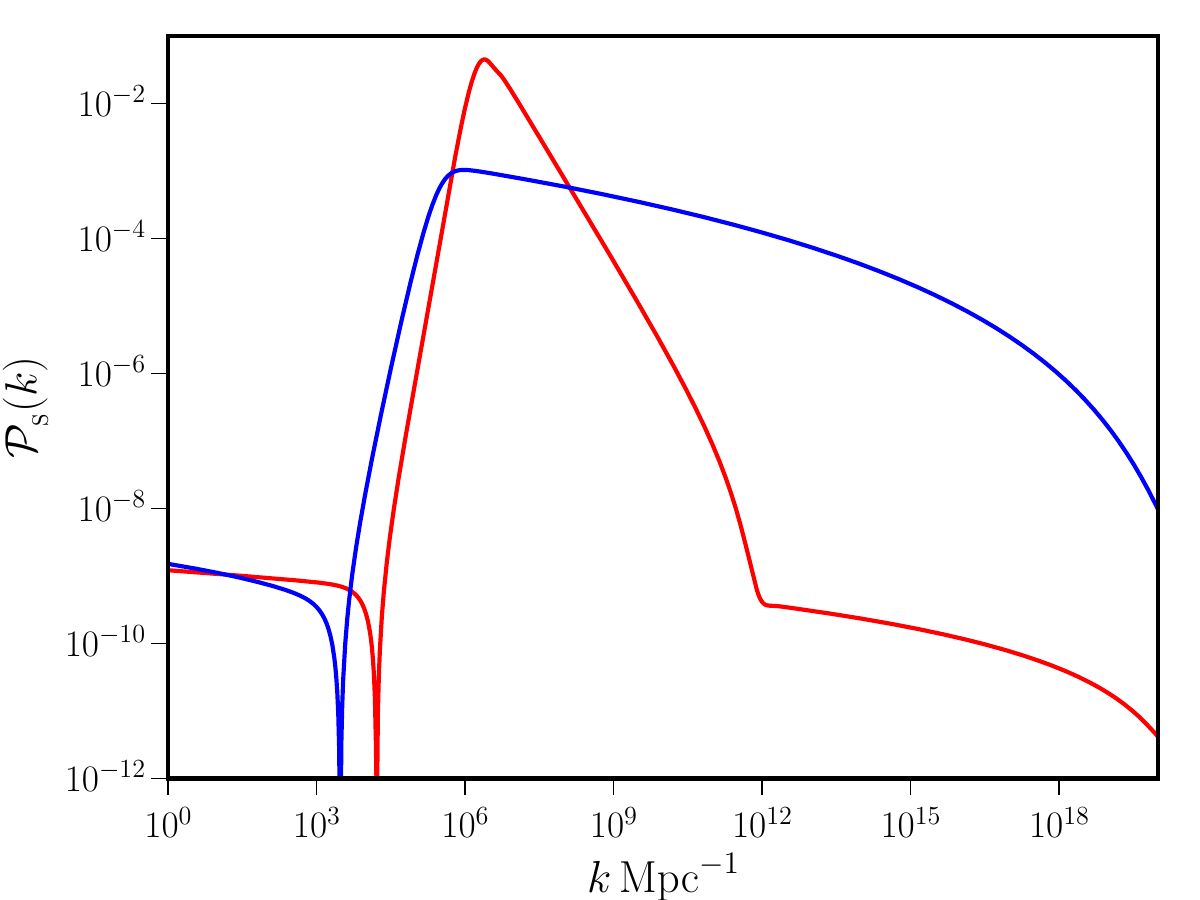}
\includegraphics[width=7.5cm, height=5cm]{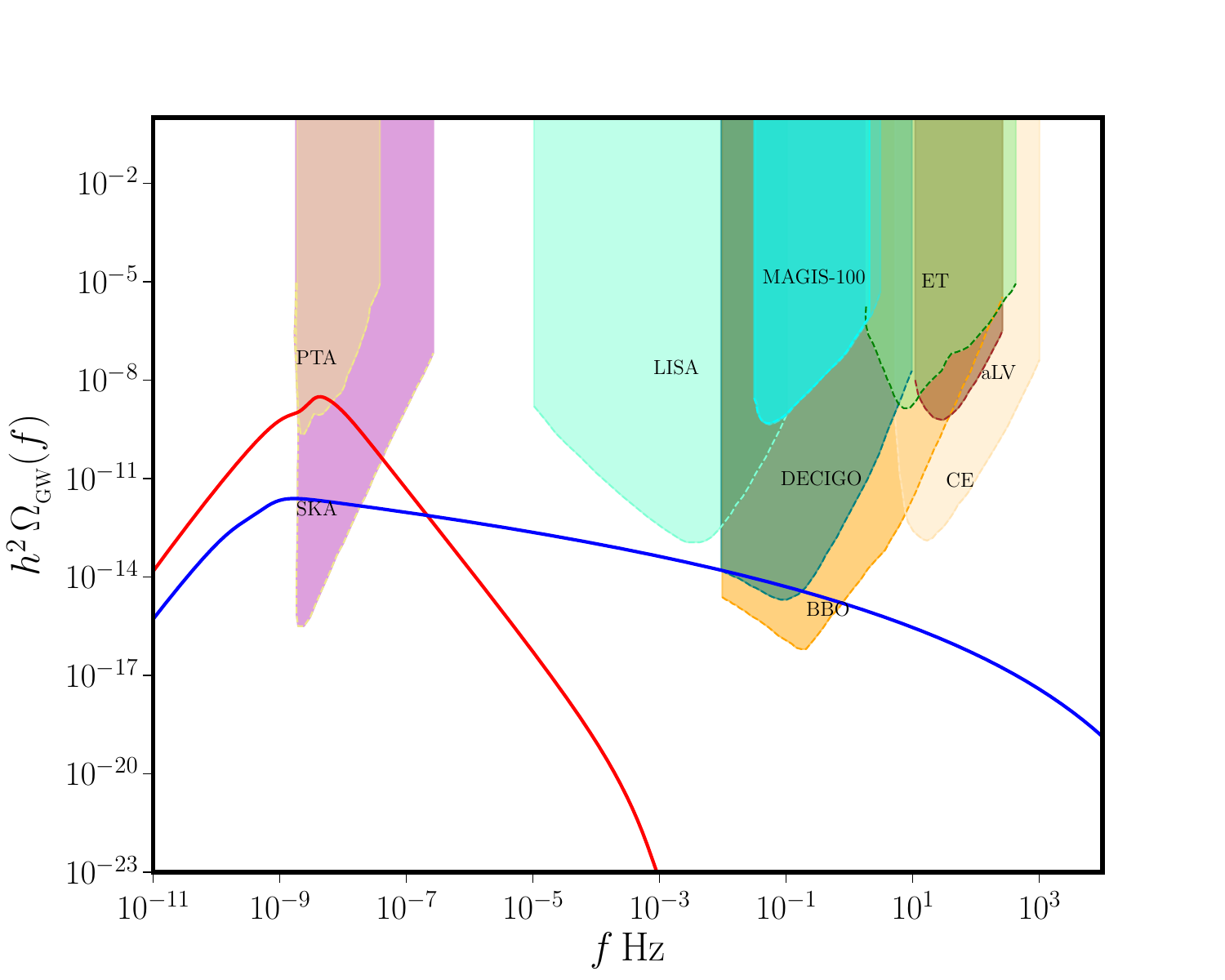}
\caption{The scalar power spectra (on the left) and the corresponding $\ogw$ 
generated (on the right) in the models of interest are presented here 
(SMD in red and CHI in blue). 
For the values of parameters chosen for these models, we observe that the 
peaks of these spectra occur at around $10^6\,\mpcinv$. This leads to the 
maximum amplitudes of associated $\ogw$ occur at around $10^{-9}\,{\rm Hz}$. 
The various constraint and sensitivity curves corresponding to current and 
upcoming GW missions are presented as shaded regions of different colors at 
the top of the plot of $\ogw$ (on the right). The intersection of $\ogw$ curve 
of CHI with the sensitivity regions of SKA and BBO indicate predictions for the 
corresponding future detectors. 
The intersection of the $\ogw$ curve of SMD with PTA indicates the possibility 
of arriving at constraints on the associated model parameters by comparing 
with the NANOGrav data~\cite{NANOGrav:2020bcs}.}
\label{fig:ps-sgw}
\end{figure}

The scalar power spectra arrived at from these models are presented in the left
panel of Fig.~\ref{fig:ps-sgw}. The power over small scales have been amplified
by several orders due to the ultra slow roll epochs in these models.
The parameters that we have worked with ensure that the spectra are COBE 
normalized over the CMB scales. However, we should mention that the predictions 
of $\ns$ and $r$ over these scales have some tension with the constraints on 
these parameters arrived at by Planck~\cite{Akrami:2018odb}. This issue is known 
in case of models with enhancement of power over small scales and the tension 
with data is larger if the peak is closer to CMB 
scales~\cite{Bhaumik:2019tvl,Ragavendra:2020sop}.
Moreover, the rise in power occurs close to the range of scales that can be
probed by the effect of spectral distortion in 
CMB~\cite{Chluba:2012we,Jeong:2014gna,Kite:2020uix}.
Hence there is a possibility of constraining these models against data from 
future missions probing this effect with improved 
sensitivity~\cite{Chluba:2019nxa}.
In this work, we shall focus on the generation of secondary GWs due to the rise 
in power over small scales and the contributions due to scalar bispectrum.

We first compute the amplitude and behavior of secondary GWs generated from 
these two models due to Gaussian contribution.
We present the observable quantity of interest, \viz 
the dimensionless energy density of secondary GWs, $\ogw$ as a function 
of frequency $f$. The spectrum of $\ogw(f)$ has been plotted for
our models of interest in the right panel of Fig.~\ref{fig:ps-sgw}. 
The peak in these spectra occur at around $10^6\,\mpcinv$ for the choices of 
parameter values we have worked with. The peak produced in the case of 
SMD is sharper than in the model of CHI. 
We also plot the constraint and sensitivity curves from various current 
and upcoming observational missions (see Ref.~\cite{Moore:2014lga} and the 
associated web-page for the sensitivity curves of various missions). 
We find that the maximum amplitude of the $\ogw$ generated is over the range 
corresponding to PTA and SKA surveys and the curve due to SMD already 
intersects with the PTA constraint. 
This indicates possible constraining and ruling out of regions in the 
parameter space determining the dip in the potential using the NANOGrav 
data~\cite{Arzoumanian:2018saf,NANOGrav:2020bcs}.

Our primary objective in this work is to examine the possible imprints of the 
scalar non-Gaussianity on the power spectra and on the $\ogw(f)$ 
in these models. Hence, we begin by calculating the correction to the power 
spectrum by the procedure discussed earlier in section~\ref{sec:pc}. 
We first compute the scalar bispectrum for the models. We evaluate all the 
contributions arising from the third order action governing the scalar 
perturbations and arrive at the complete form of the scalar bispectra 
$G(k_1,k_2,k_3)$ [\cf Eqs.~\eqref{eq:sbs} and \eqref{eq:cG}] 
for each of the models. We then use the relation given in 
Eq.~\eqref{eq:fnl-work}, to obtain the associated $\fnl(k_1,k_2,k_3)$. 
This $\fnl(k_1,k_2,k_3)$ is then substituted into Eq.~\eqref{eq:pc-fnl}, 
to finally arrive at the correction to the power spectrum $\pc(k)$. 
Since the bispectra for the models of interest are not easy to evaluate 
analytically, we perform this calculation numerically.

\begin{figure}[!t]
\includegraphics[width=7.5cm, height=5cm]{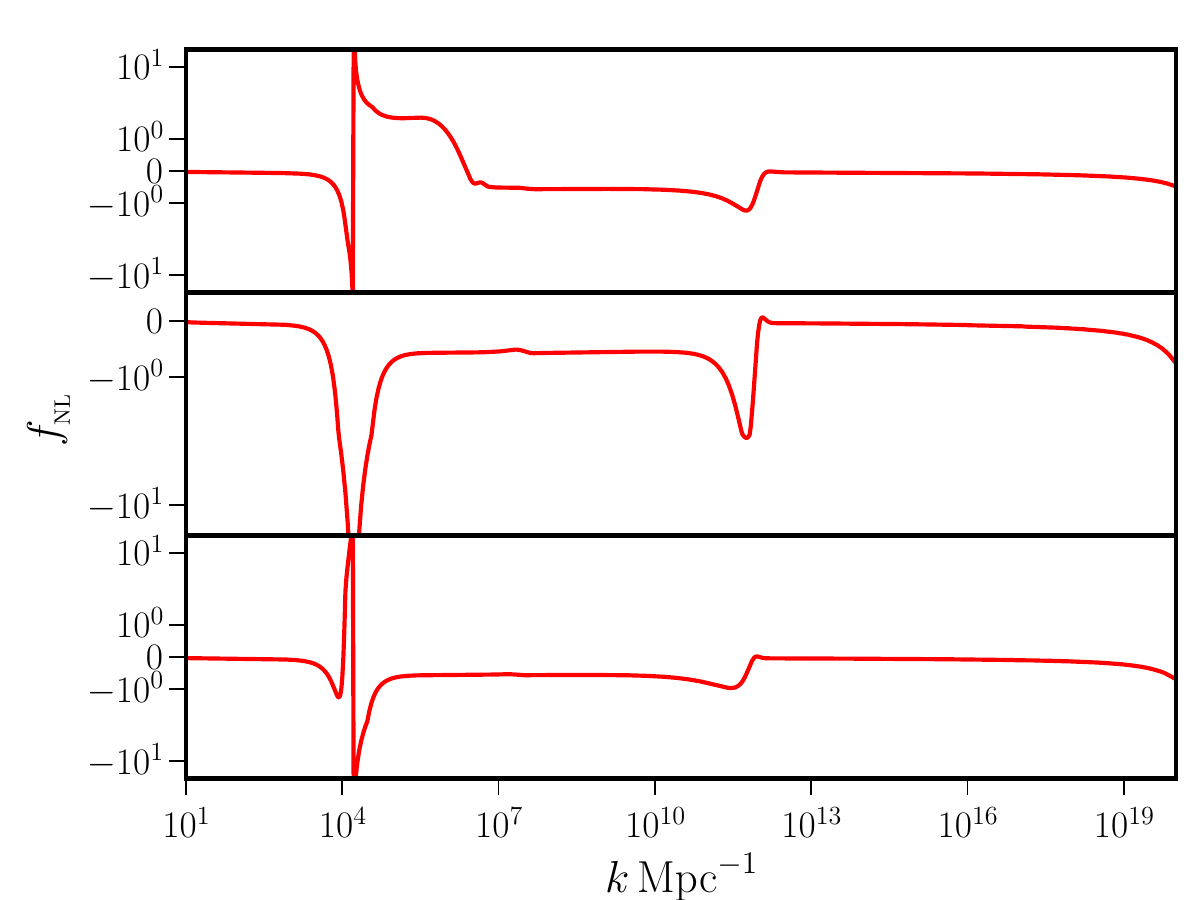}
\includegraphics[width=7.5cm, height=5cm]{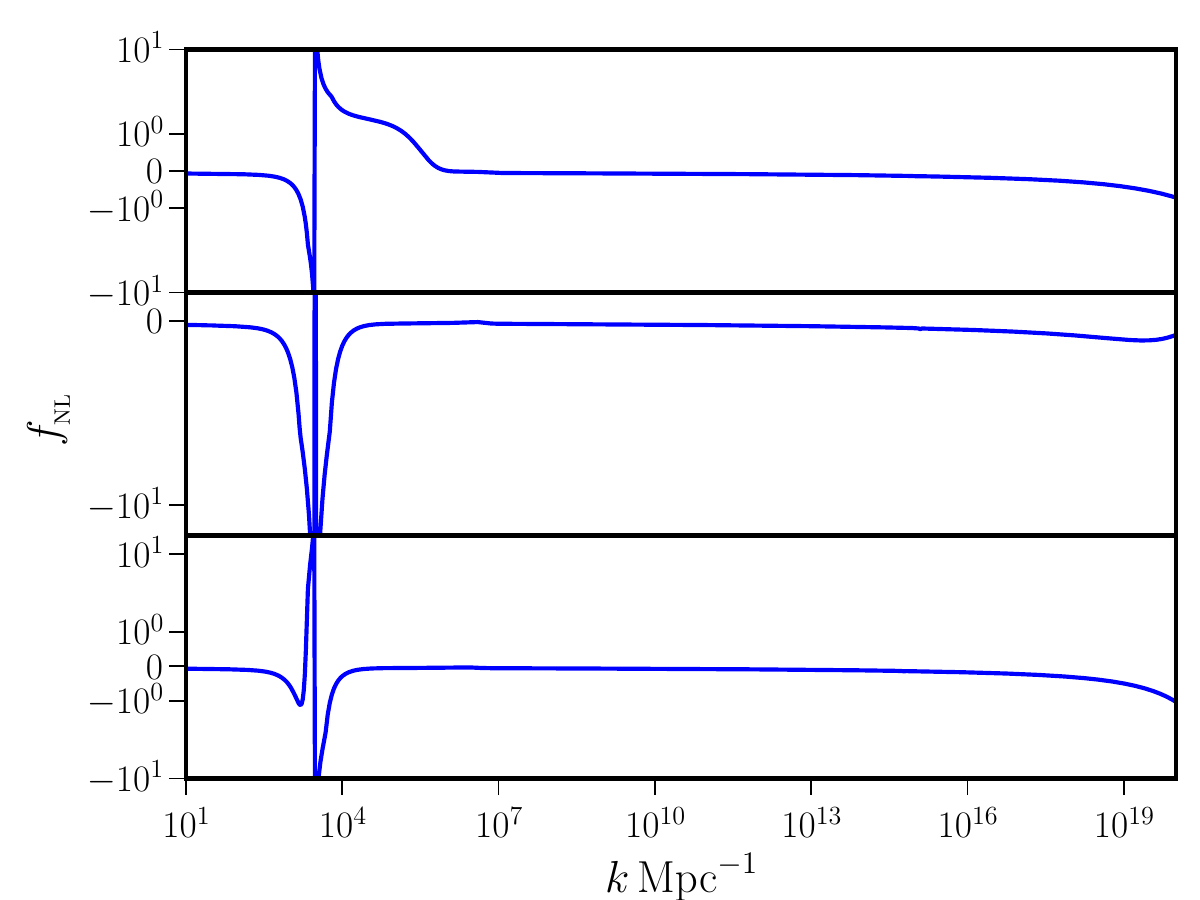}
\caption{We present the non-Gaussianity parameter $\fnl(k_1,k_2,k_3)$ for the
two models of interest (SMD in left and CHI in right) in various limits,
\viz squeezed limit (on the top), equilateral limit (in the middle) and 
the flattened limit (in the bottom panel). 
We see that the $\fnl(k_1,k_2,k_3)$ has non-trivial scale dependence and it is 
important to capture its complete behavior while computing the corrections to 
the power spectrum. 
There are rather large values of $\fnl(k_1,k_2,k_3)$ occurring at the 
wavenumbers corresponding to the location of the sharp downward spike in the 
power spectra of respective models. As mentioned earlier, these spuriously 
large values should be dealt with caution and have to be regulated while using 
$\fnl(k_1,k_2,k_3)$ in further calculations.}\label{fig:fnl-compare}
\end{figure}

\begin{figure}[!t]
\includegraphics[height=5.0cm, width=8.5cm]{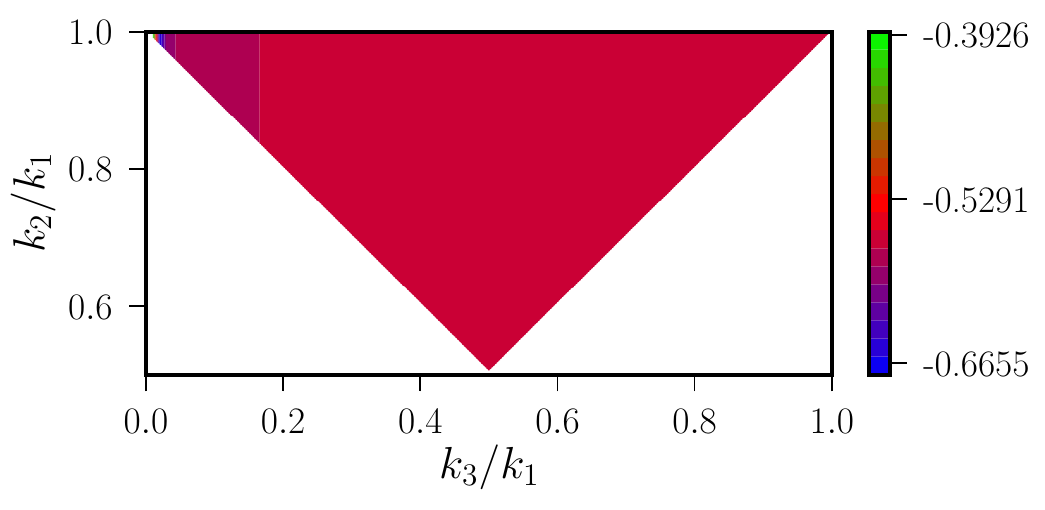}
\includegraphics[height=5.0cm, width=8.5cm]{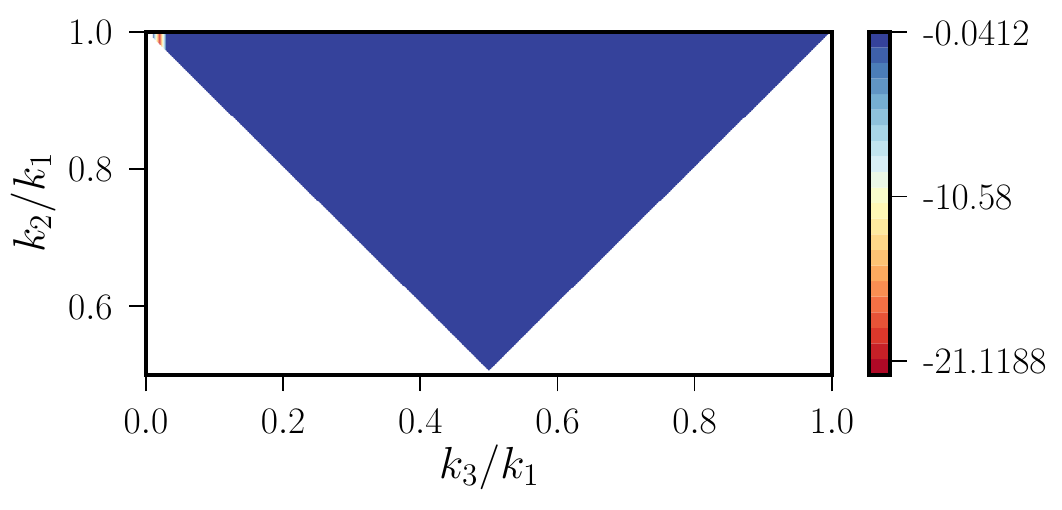}
\caption{The density plots of the scalar non-Gaussianity parameter 
$\fnl(k_1,k_2,k_3)$ illustrating its behavior in a general configuration of
wavenumbers around a given value of $k_1$ is presented for models of interest. 
The parameter obtained from SMD is plotted on the left, while the parameter 
from the model of CHI on the right. 
The behavior is evidently dependent on the value of $k_1$, which for both the
models is taken to be $k_1=10^6\,\mpcinv$, corresponding to the wavenumber
close to the peak in the spectra.
We find that the $\fnl(k_1,k_2,k_3)$ in these models are highly local in shape
just around the peak in the spectra. The value of the parameter is roughly $-0.5$
in case of SMD, whereas in case of CHI, it turns out to be around $-0.04$.
As we move away from the peak, with values of $k_3 \ll k_1$, we see that
the $\fnl$ starts deviating from the local shape and growing larger in value.}
\label{fig:fnl-2D-compare}
\end{figure}

We present the behavior of this parameter, in Fig.~\ref{fig:fnl-compare}, 
for both the models of interest in various limits of the configuration of 
wavenumbers, \viz, 
the squeezed limit ($\vkc \to {\bf 0},\,\vka = -\vkb$), 
equilateral limit ($k_1 = k_2 = k_3 = k$) and the 
flattened limit ($k_1 = k_2 = k,\,k_3 = 2\,k$). 
The parameter exhibits non-trivial behavior close to the wavenumber 
corresponding to the peak in the power spectra. The behavior is smoother over 
scales farther from the peak in the spectra.
We also present the density plot of $\fnl$ around the peak in the power for
these models in Fig~\ref{fig:fnl-2D-compare}. We find that $\fnl$ is largely
local in its behavior around the peak. We should note that the value of $\fnl$
is lesser than unity over this range of wavenumbers close to the peak. 
However, we notice deviation from these local values as we move further from 
the peak, \ie $k_3$ takes values smaller than $k_1$.
We should mention here that there arises a sharp spike in the 
$\fnl(k_1,k_2,k_3)$ at the point where there is a sharp downward spike in the 
power spectrum, occurring before the rise and the peak in the range of 
wavenumbers. This indicates power spectrum reaching very small values.
Hence, quantities such as $\fnl(k_1,k_2,k_3)$ or the scalar spectral index
$\ns(k)$ that contain power spectrum in their denominators of their 
definitions, may incur spuriously large values at this wavenumber. 
Therefore, care should be taken when dealing with such anomalous values. 
In our calculation, we have regulated the value of $\fnl$ around the region 
by introducing a cutoff of $10$. This implies that any value of $|\fnl|$ which 
is larger than $10$ is taken to be $10$.

\subsection{Calculation of the correction}
With $\fnl(k_1,k_2,k_3)$ thus computed, we can obtain the correction to the 
spectrum $\pc(k)$ for both the models. Before we proceed to perform the 
integrals numerically, we consider the Eq.~\eqref{eq:pc-fnl} and attempt to 
arrive at a rough analytical estimate of the $\pc(k)$.

\begin{figure}[!t]
\includegraphics[height=5.0cm, width=7.5cm]{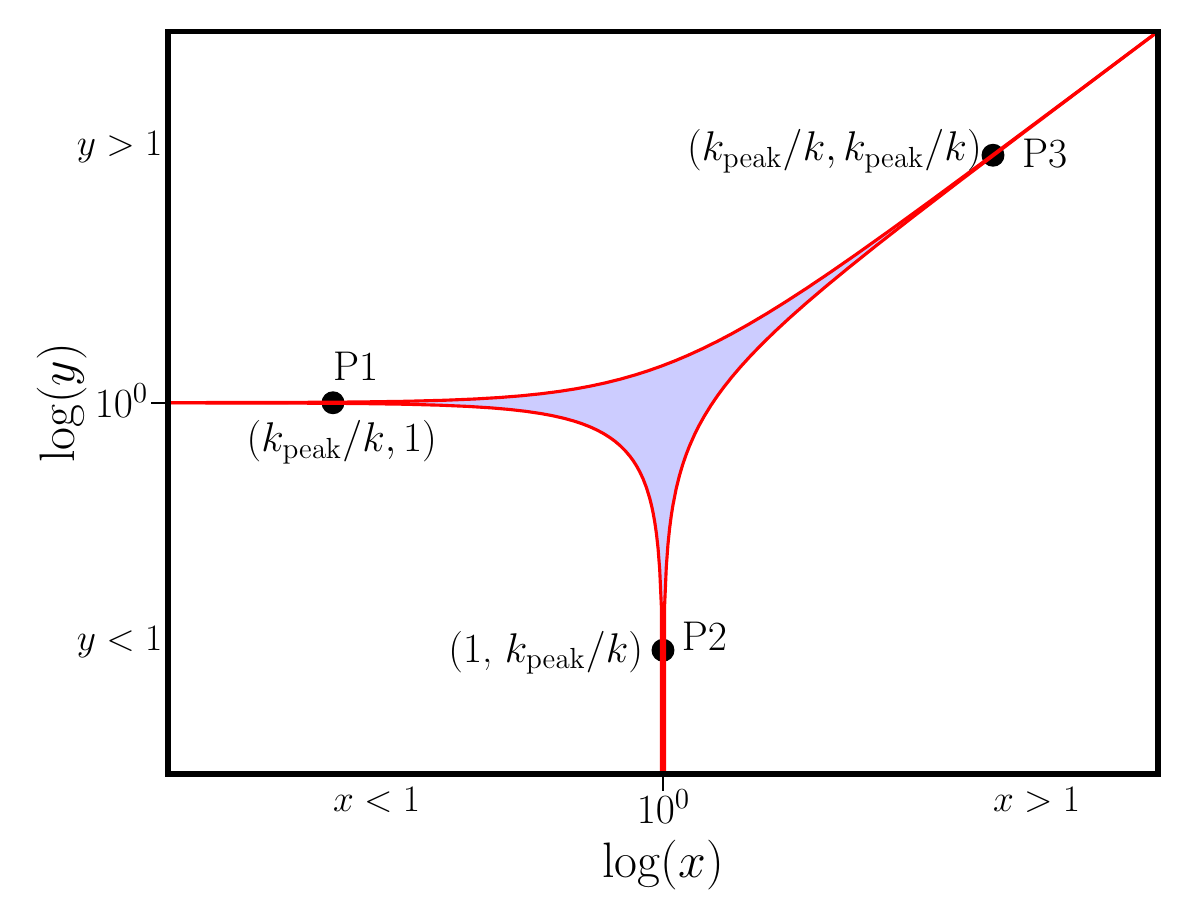}
\caption{The range of integration involved in calculating $\pc(k)$ 
[\cf Eq.~\eqref{eq:pc-fnl}] is plotted in logarithmic scale. The shaded region 
marks the region covered by the limits of the integrals. We mark the three 
points P1, P2 and P3 at which the integrals derive maximum contribution 
when there is a localized peak in the power spectrum. The region around the 
points P1 and P2 contribute for $k > \kf$ and the region around P3 contributes 
for $k < \kf$. It is also worth noting that, due to the symmetry of the 
integrand over the variables $x$ and $y$ the contributions from P1 and P2 
turn out to be equal to one another. 
For the case of $k \sim \kf$ the integrand receives the maximum contribution 
from the wide region around $x=y=1$.}
\label{fig:integrand}
\end{figure}

Let $\kf$ denote the wavenumber corresponding to the peak in the power spectrum.
We know that the maximum amplitude of the integrand occurs around the region
where $x=\kf/k$ or $y=\kf/k$ or $x=y=\kf/k$. We illustrate the range of the 
integrals involved and the points where the maximum contribution arises from
in Fig.~\ref{fig:integrand}. We shall describe the sharp peaking behavior of 
the power spectrum by approximating its form around the peak using a Dirac 
delta function as 
\begin{eqnarray}
\ps(k) = \ps(\kf)\,\delta(\ln(k)-\ln(\kf))\,.
\label{eq:ps-dirac}
\end{eqnarray}
Using this approximation, we proceed to compute the dominant contributions to 
the integrals. We perform the integral over $x$ in Eq.~\eqref{eq:pc-fnl} 
to obtain that
\begin{eqnarray}
\pc(k) &=& \f{9}{25}\bigg(\f{k}{\kf}\bigg)\ps(\kf)
\int^{1+\kf/k}_{\vert 1-\kf/k \vert}
\f{\d y}{y^2}\ps(ky)\,\fnl^2(k,\kf ,ky)
\end{eqnarray}

Now we shall consider the two regimes in wavenumbers, \viz $k < \kf$ and 
$k > \kf$. For the case of $ k < \kf$, the integrand receives contribution only 
from the point P3 marked in Fig.~\ref{fig:integrand}.
Due to the narrow range of the integral over $y$, we may approximate 
$ky \simeq \kf$ in the arguments of $\ps$ and $\fnl$. So, the
above integral over $y$ simplifies to
\begin{eqnarray}
\pc(k) &\simeq& \f{9}{25}\bigg(\f{k}{\kf}\bigg)
\left[\ps(\kf)\,\fnl(k,\kf,\kf)\right]^2
\int^{1+\kf/k}_{\vert 1-\kf/k \vert}\,
\f{\d y}{y^2}\, \nn \\
 &=& \f{18}{25}\bigg(\f{k}{\kf}\bigg)^3\left[\ps(\kf)\fnl(k,\kf,\kf)\right]^2\,,
\end{eqnarray}
where we have used the fact that $\kf/k > 1$. It is interesting to note the 
combination of wavenumbers appearing in the argument of $\fnl$. We know that
$k < \kf$. Hence, $\fnl(k,\kf,\kf)$ denotes that the parameter has to be
evaluated in the squeezed limit of the configuration of wavenumbers. 
This further simplifies the expression because we know that the consistency 
condition relating the $\fnl$ and the scalar spectral index $\ns(k)$ is 
obeyed in these models~\cite{Ragavendra:2020sop,Zhang:2021vak}. Therefore, 
we utilize the consistency relation \ie
\begin{eqnarray}
\fnl(k,\kf,\kf) &=& \f{5}{12}[\ns(\kf)-1]\,.
\end{eqnarray}
Here, strictly speaking, $[\ns(\kf)-1]$ vanishes identically since it is the
slope of the spectrum at its peak. However, we shall take it to be a small
non-vanishing value close to the peak in the spectrum for the purpose of our 
calculation. Therefore expression for $\pc(k)$ reduces to
\begin{eqnarray}
\pc(k) &=& \f{1}{8}\bigg(\f{k}{\kf}\bigg)^3
\left\{\ps(\kf)\,[\ns(\kf)-1]\right\}^2\,.
\end{eqnarray}
We find that $\pc(k)$ shall be proportional to $k^3$ over the scales with
$k < \kf$.

We then consider the case of $k > \kf$. For these wavenumbers, there arise 
contributions from two points, P1 and P2 as marked in Fig.~\ref{fig:integrand}.
We shall first evaluate the contribution at P1 using the approximation of 
the spectrum in Eq.~\eqref{eq:ps-dirac}. The expression for $\pc(k)$ becomes
\begin{eqnarray}
\pc(k) &\simeq& \f{9}{25}\bigg(\f{k}{\kf}\bigg)\,\ps(\kf)\ps(k)
\fnl^2(k,\kf, k)\int^{1+\kf/k}_{\vert 1-\kf/k \vert}
\f{\d y}{y^2}\, \nn \\
 &=& \f{18}{25}\,\ps(\kf)\ps(k)\fnl^2(k,k,\kf)\,,
\end{eqnarray}
where we have used the smallness of $\kf/k$. We again note that the arguments
of $\fnl$ suggest that it is evaluated in the squeezed limit but now with 
$\kf$ acting as the squeezed mode. Hence we shall make use of the 
consistency relation again, where
\begin{eqnarray}
\fnl(k,k,\kf) &=& \f{5}{12}[\ns(k)-1]\,.
\end{eqnarray}
This reduces the expression for $\pc(k)$ to
\begin{eqnarray}
\pc(k) &=& \f{1}{8}\ps(\kf)\ps(k)\left[\ns(k)-1\right]^2\,.
\end{eqnarray}
Due to the fact that the form of the integral in Eq.~\eqref{eq:pc-fnl} 
remains unchanged under the exchange of $x$ and $y$, the contribution
from the point P2 shall be the same as given above. So, we have the
total value of $\pc(k)$ for $k > \kf$ to be
\begin{eqnarray}
\pc(k) &=& \f{1}{4}\ps(\kf)\ps(k)\left[\ns(k)-1\right]^2\,.
\end{eqnarray}
We find that $\pc(k)$ over this regime of $k > \kf$ shall be 
proportional to $\ps(k)$ with no explicit scale dependence. If the spectrum
turns nearly scale invariant away from the peak over large wavenumbers, then 
we can expect a corresponding $\pc(k)$ with nearly constant amplitude.
In summary, we have the analytical estimate of $\pc(k)$ to be
\begin{equation}
\pc(k) = \begin{cases}
~\displaystyle\f{1}{8}\bigg(\f{k}{\kf}\bigg)^3 
\left\{\ps(\kf)[\ns(\kf)-1]\right\}^2\,, & ~~\text{for}\,k<\kf\,, \\
~\displaystyle\f{1}{4}\,\ps(\kf)\ps(k)
\left[\ns(k)-1\right]^2\,, & ~~\text{for}\,k>\kf\,.
\end{cases}
\end{equation}

Having obtained these analytical expressions, we proceed to compute the 
exact numerical estimates of $\pc(k)$. We shall briefly discuss certain aspects
of numerical evaluation of the integrals involved.
The integral is evaluated ensuring that the regime of $x = \kf/k$ and 
$y=\kf/k$ are well sampled. Due to the wide range of the integral over 
$x$, the integration is performed over log scale. The limits are chosen such 
that the range of integration is centered at $\kf/k$ and spans two decades on 
either side of the point. For given values of $kx$ and $ky$, the power spectra 
is evaluated numerically.
Besides, each point of this $x$--$y$ plane provides a triangular configuration 
of wavenumbers for which $\fnl(k,kx,ky)$ is evaluated numerically. 
This is the most time consuming part of the calculation. Once computed, 
the integrand is summed over to obtain $\pc(k)$. The exercise is repeated
for complete range of wavenumbers.


\subsection{Calculation of non-Gaussian contributions to $\ogw$}

The behavior of $\pc(k)$ may give us an idea of the effect of $\fnl$ on the
scalar power spectrum. It may further give us an insight about the amplitude
of one of the non-Gaussian contributions to $\ogw$~[\cf Eqs.~\eqref{eq:ph-fnl2}]. 
Having obtained the $\pc(k)$ in the models of interest, we proceed to compute
the non-Gaussian contributions $\ph^{(2-i)}$ to the $\ogw$ in these cases.

At the outset, we should note that, the non-trivial dependence of $\fnl$ over 
different combination of wavenumbers in $\ph^{(2-1)}(k)$, $\ph^{(2-2)}$ and 
$\ph^{(2-3)}(k)$ do not allow us to easily obtain an analytical estimate as 
we did for $\pc(k)$.
Hence, as mentioned earlier, we numerically perform these integrals involved 
using the Monte-Carlo method of integration. 
Let us now mention a few details about the procedure.
We first identify the region of maximum amplitude of the integrands in the range
of integration, for a given wavenumber $k$. 
Interestingly, we find that the integrands have maximum values 
around the wavenumber $\kf$, if the wavenumber of interest $k < \kf$, 
while they peak around $k$ if $k > \kf$. 
We also find that the nature of integrands are very localized in the range 
of $k$ allowing us to set the range of integrals to be two decades on either 
side of the peaks of the integrands. 
The respective angular integrals are performed over the entire range 
\viz $\cos \theta_i \in [-1,1]$ and $\phi_i \in [0,2\,\pi]$. 
During the performance of integration, each point corresponds to numerical 
evaluation of a combination of power and bi-spectra with their appropriate 
arguments of wavenumbers. 
The computation of $\fnl$ at each point of integration is the time consuming 
part of this process. 
The integrals were performed using $10^5$ points and checked for convergence.

We should also note an interesting property of these contributions. The 
integrand describing $\ph^{(2-2)}(k)$ is positive definite and hence the 
contribution shall be positive. However, the integrand characterizing the 
contributions $\ph^{(2-1)}(k)$ and $\ph^{(2-3)}(k)$ can be negative, because 
of their dependence over the polar angles (as noted earlier in 
Ref.~\cite{Adshead:2021hnm}). This property should be accounted for while 
comparing them against $\ogw$ obtained from the Gaussian contribution. 


\subsection{Results}

\begin{figure}
\centering
\includegraphics[width=7.25cm]{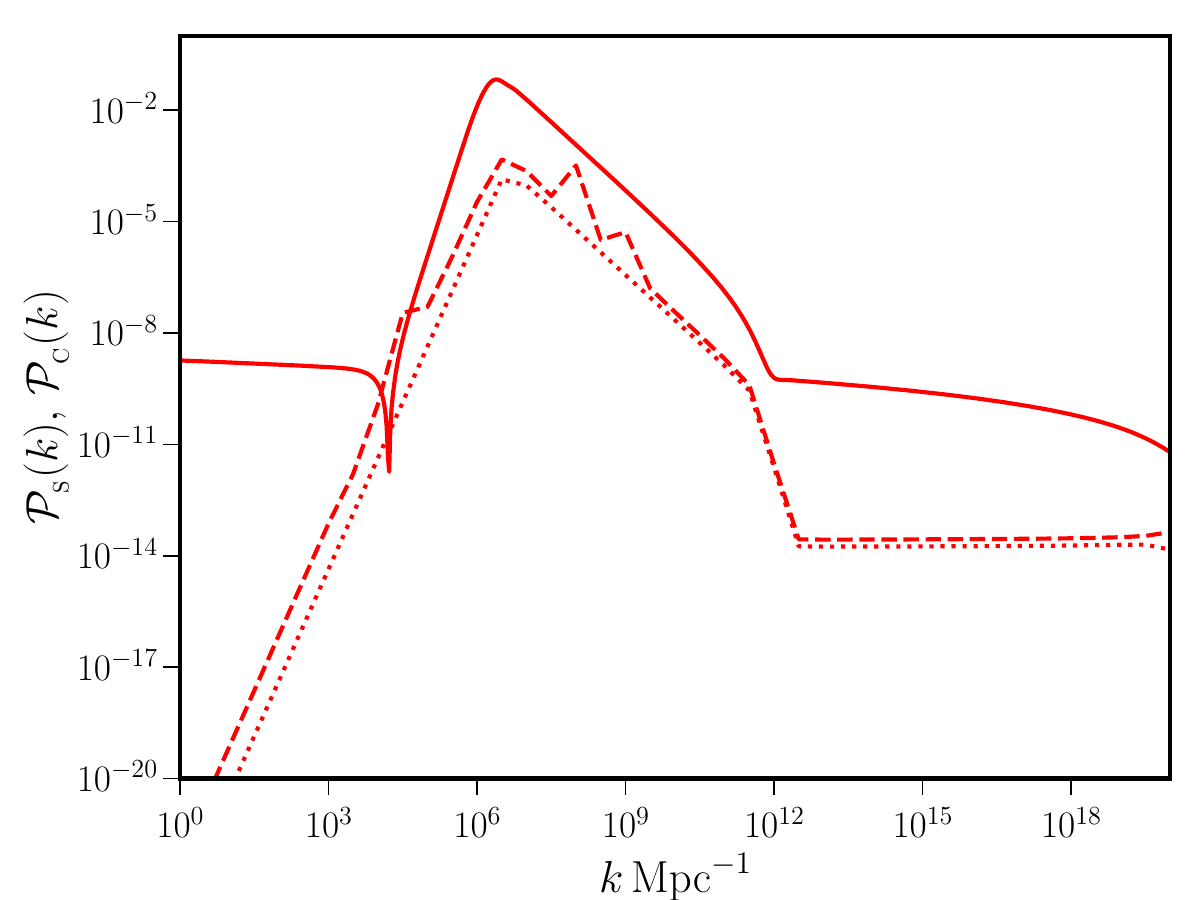}
\includegraphics[width=7.25cm]{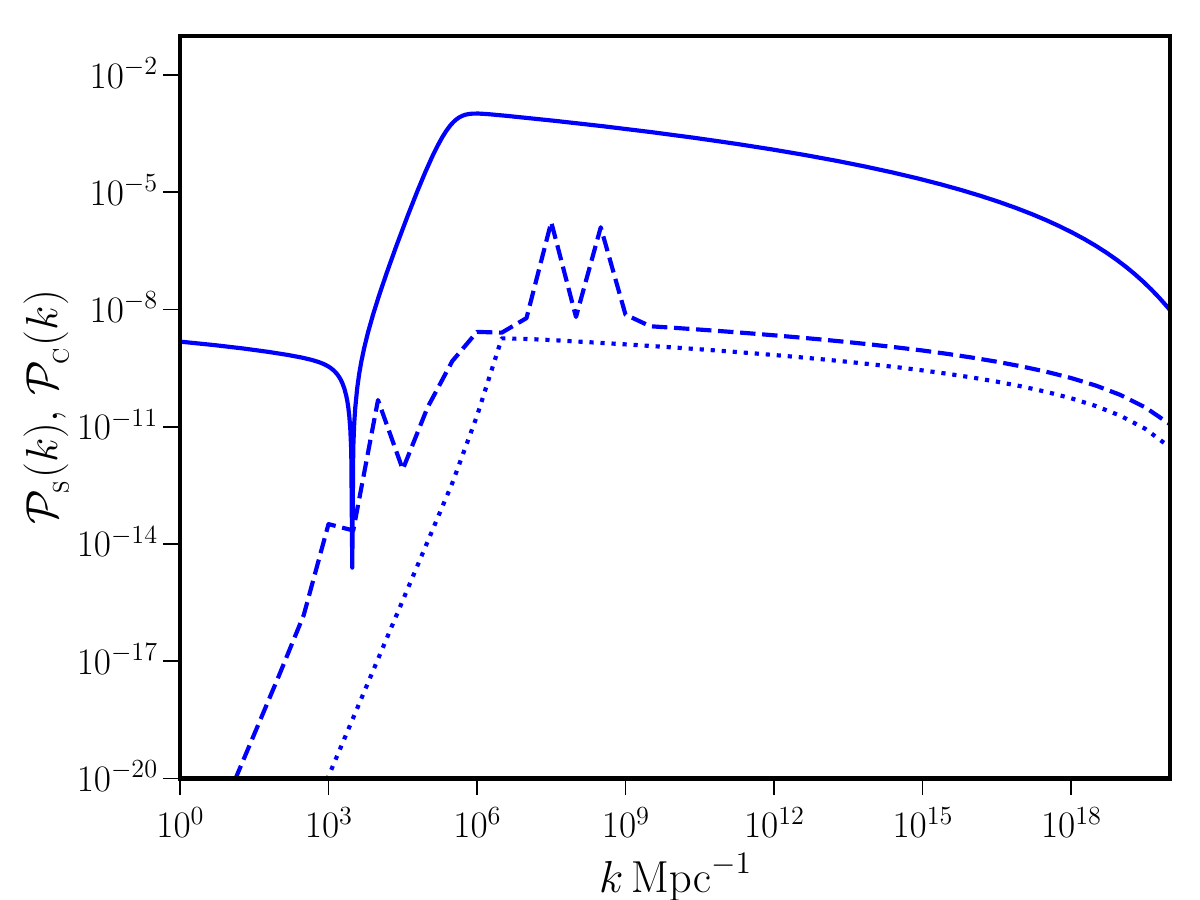}
\caption{The original scalar power spectra $\ps(k)$ (as solid 
lines) and the non-Gaussian corrections $\pc(k)$ due to the bispectrum
(as dashed lines) have been plotted here for the models of interest,
\viz SMD (on the left) and CHI (on the right). 
Evidently, the $\pc(k)$ computed is lower in amplitude than $\ps(k)$. 
We have also plotted the analytical estimate of $\pc(k)$ for these two models 
(as dotted lines). The analytical estimate matches the numerical 
behavior better in the case of SMD than CHI since the spectrum is 
more sharply peaked in the first model than in the second model. 
The complete spectrum corrected for $\pc(k)$ shall effectively be the 
same as the original $\ps(k)$, particularly around $\kf$ and over 
$k > \kf$.}\label{fig:ps-corr}
\end{figure}

First, we present the $\pc(k)$, obtained both the analytically and numerically, 
against the original spectra, $\ps(k)$, in Fig.~\ref{fig:ps-corr}. 
We observe that $\pc(k)$ is smaller than the original $\ps(k)$ particularly
around the peak and over the range $k > \kf$. 
There appears a region close to the dip in the spectrum where 
$\pc(k)$ is greater than $\ps(k)$. This is mainly due to the sharp
spike occurring in $\fnl$ that we mentioned earlier. But apart from
this effect, there arises no significant correction to the original
power spectrum. Moreover, the analytical estimate fairly mimics the exact 
numerical behavior of $\pc(k)$. The behavior of $k^3$ over large scales and 
near scale invariance over small scales is well captured in the numerical 
result thereby assuring the validity of the analytical estimates over
wavenumbers far from the peak.
The match is better for the model SMD. This can be understood because its 
spectrum is closer in resemblance to the Dirac delta function used in the 
analytical calculation. The original spectrum $\ps(k)$ in case of CHI has a
rather broad peak with slower descent over the range of wavenumbers. 
This behavior leads to the difference between numerical and analytical estimates
of $\pc(k)$ around the peak in this model. However, for $k > \kf$, the 
analytical estimate matches better even in case of such a broad peak.
The rugged nature of the numerical result is due to the limited number of 
points taken for evaluation over the range of wavenumbers.

\begin{figure}
\includegraphics[width=7.25cm]{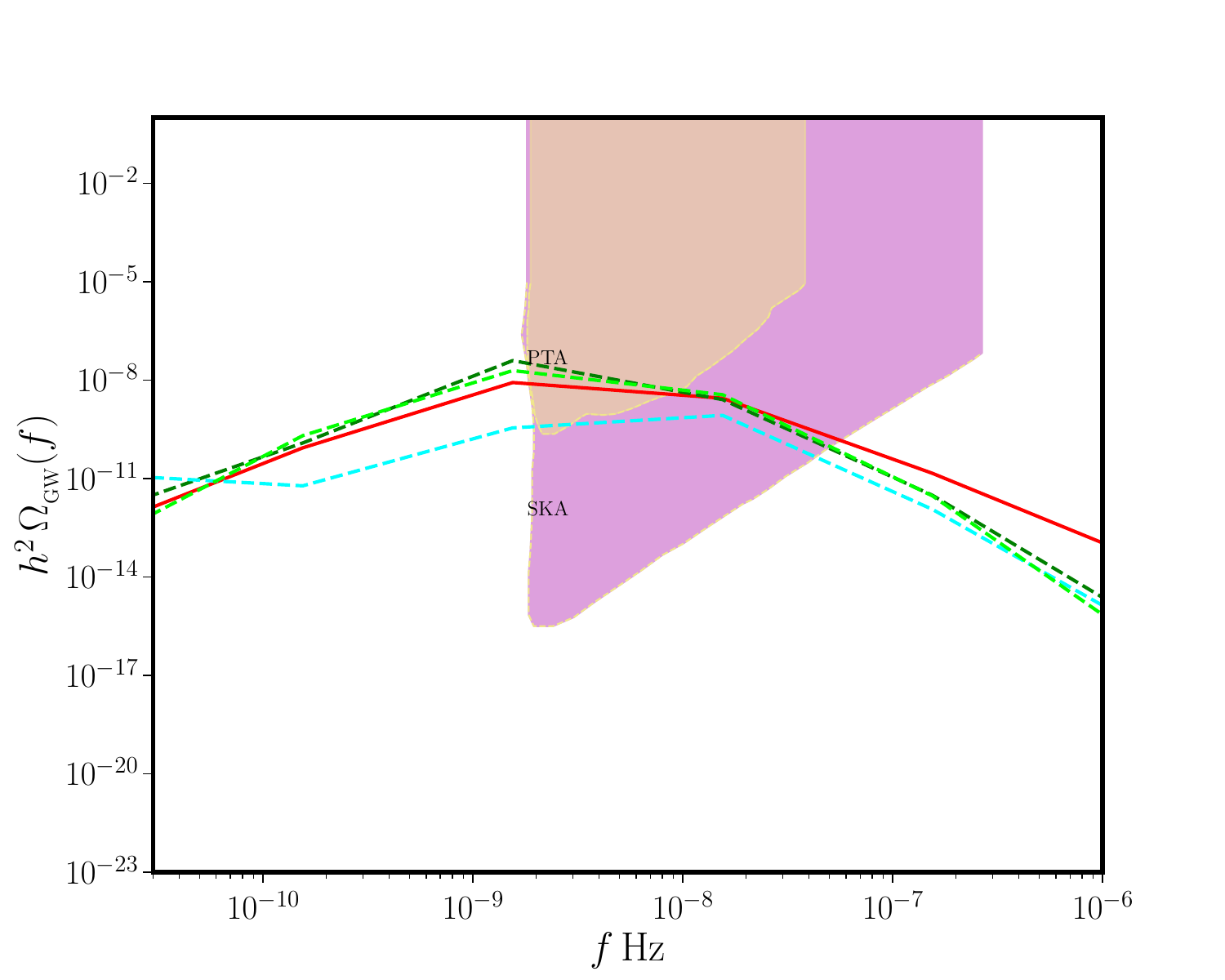}
\includegraphics[width=7.25cm]{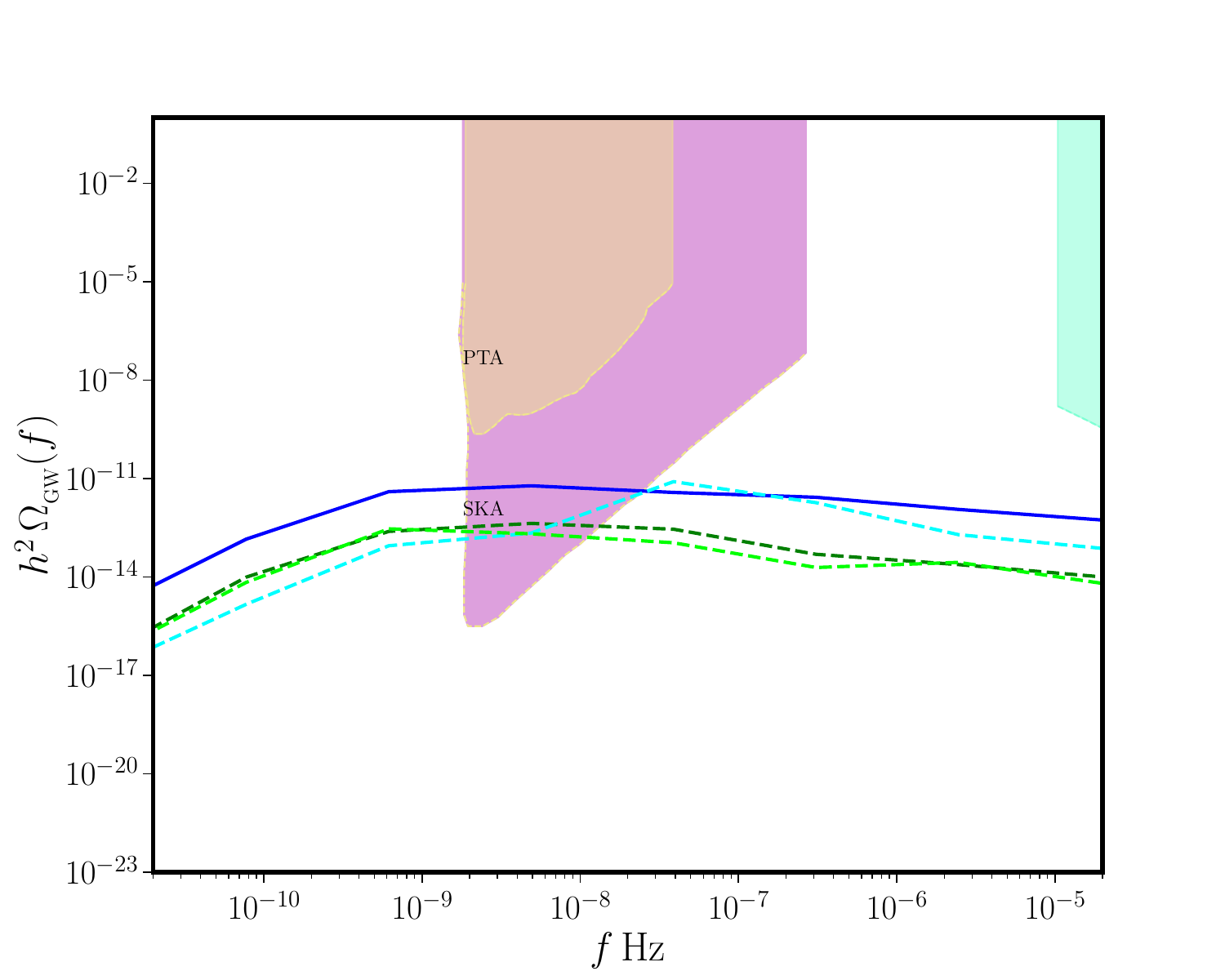}
\caption{We present the non-Gaussian contributions to $\ogw$ arising due to 
$\fnl$ (as dashed lines), against the original Gaussian contribution 
(as solid lines) for the models SMD (on left) and CHI (on right), focusing 
over the range of frequencies containing the maximum amplitude.
The contributions arising from the terms $\ph^{(2-1)}(k)$ (in green), 
$\ph^{(2-2)}(k)$ (in cyan), $\ph^{(2-3)}(k)$ (in lime) are presented for both 
the models of interest.}\label{fig:sgw-corr}
\end{figure}

We present the behavior of the non-Gaussian contributions to $\ogw$ at the 
level of $\fnl^2$, \viz due to $\ph^{(2-i)}(k)$, in Fig.~\ref{fig:sgw-corr}. 
We focus particularly around the peak amplitude of $\ogw$ and find that the 
non-Gaussian contributions are significant for SMD. These contributions 
dominate the $\ogw$ from the Gaussian spectrum for wavenumbers $k \leq \kf$.
However, they become sub-dominant for $k>\kf$. In case of CHI, the 
non-Gaussian contributions become briefly comparable over the range of 
$k \simeq \kf$. But they are sub-dominant for wavenumbers $k<\kf$ as well 
as $k>\kf$. Thus, we learn that the behavior of $\ogw$ arising from 
non-Gaussian contributions are highly model dependent and significant in case of
power spectrum with highly localized behavior around the peak. 
However, as we move farther from the peak, these contributions become lesser 
in amplitude compared to the Gaussian contribution. 
Therefore, these models illustrate that the non-Gaussian contributions to 
$\ogw$ have to be computed and consistently accounted for, especially around 
the peak of the spectral density.


\section{Conclusion}\label{sec:conc}

There have been attempts in the literature to account for scalar 
non-Gaussianity in the calculation of the spectral density of
the secondary GWs, $\ogw(f)$, for specific cases of $\fnl$ 
assuming certain shapes or limits of the bispectrum.
In this work, we have presented a method to account for a general scalar 
bispectrum with non-trivial scale dependence in such a calculation. 
We have presented the correction to the scalar power spectrum that may arise 
due to the scalar bispectrum. We have also attempted an analytical estimate
of the correction to be expected from models with a localized peak in the 
power spectrum. We have found that it is the squeezed limit of $\fnl$ that 
contributes the most to the correction for wavenumbers away from the peak in 
the power spectrum.
We have then presented the non-Gaussian contributions to $\ogw(f)$ that 
arise due to $\fnl$. We have computed terms that are reducible in terms of
$\pc(k)$ as well as those that are not reducible so.
We have consistently accounted for the scale dependence of $\fnl$, arising 
from the modified definition of the parameter, in computing these 
contributions.

We then illustrated our method using two models of inflation. 
These are models driven by canonical scalar fields that permit brief epochs of 
ultra slow roll and hence lead to significant amplitudes of secondary GWs. 
We have computed the correction to the power spectrum arising from $\fnl$ and 
find that it is largely sub-dominant to the original power spectrum.
Moreover, the analytical estimate of the correction agrees fairly well with
the exact numerical estimate in these cases.
We have then computed the non-Gaussian contributions to the $\ogw(f)$ and
compared them against the Gaussian contribution. We have computed these 
contributions up to the level of $\fnl^2$. We have found that the 
non-Gaussian contributions are non-trivial and slightly different from the 
shape of the original $\ogw(f)$. 
The non-Gaussian contributions arising in the case of SMD have been found 
to dominate the original amplitudes of $\ogw(f)$ around the frequencies 
corresponding to the wavenumber $\kf$ containing the peak in the power spectrum, 
as well as smaller wavenumbers, \ie over $k<\kf$. But, these contributions 
decrease farther from the peak and become sub-dominant to the Gaussian
contribution for wavenumbers with $k > \kf$. In the case of CHI, the 
non-Gaussian contribution become briefly comparable to and dominant over the 
Gaussian contribution around $\kf$, but remain sub-dominant farther from 
$\kf$ on either side.
Since the models serve as examples typical models of 
inflation that are considered in this context of generation of secondary
GWs, we can argue that the non-Gaussian contributions arising from
$\fnl$ may turn out to be significant, particularly around the peak 
amplitude of $\ogw$. Hence, they have to be computed and accounted for in the
estimates of $\ogw$.

Besides, we should emphasize that the method used for calculation has its 
value in being able to capture the complete behavior of $\fnl(k_1,k_2,k_3)$ 
in any non-trivial scenario of inflation. 
Moreover, the analytical estimate of the correction to the power spectrum,
$\pc(k)$, serves as a good approximation for the exact estimate, without 
directly computing the bispectrum. 
This greatly reduces the time taken for the calculation of $\fnl$ and 
provides a quick estimate of $\pc(k)$ to be expected from just the shape of
the spectrum for any model with a peak in its scalar power.
Importantly, the non-negligible levels of non-Gaussian contributions to 
$\ogw$ obtained in these models indicate the necessity to capture the 
exact scale dependence of $\fnl$ as presented in this method.

As to the caveats of this work, we should mention that we have restricted 
the computation of non-Gaussian contributions to $\ogw$ up to terms 
involving $\fnl^2$ due to the complexity of numerical implementation.
We are currently working on addressing the complexity and accounting
for terms involving $\fnl^4$ in the calculation.
Secondly, there arises a spike like behavior in the shape of $\fnl$ 
[\cf Fig.~\ref{fig:fnl-compare}]. This occurs due to the presence of $\ps(k)$ 
in the denominator of the expression of $\fnl$ in terms of power and 
bi-spectra [\cf Eq.~\eqref{eq:fnl-work}]. As $\ps(k)$ reaches extremely 
small values, this spike occurs and it has to be regulated to a finite value 
during the computation. This has an effect in our results as one may see a 
corrugated shape of $\pc(k)$ computed using $\fnl$.
Hence, to avoid such artefacts in computation, it may be preferable to 
modify this method to utilize the bispectrum directly in the calculation 
of $\pc(k)$ as well as the non-Gaussian contributions to $\ogw$. We are 
presently working on these issues.

In summary, we argue that the method we have discussed is a robust way to 
account for the exact form of primordial scalar non-Gaussianity at the level 
of three point correlation in the calculation of $\ogw$ arising from models of 
inflation. 
Since we infer a significant non-Gaussian contribution to
$\ogw$ in the models considered, it would be interesting to employ this 
method for non-canonical models that can potentially produce larger 
amplitudes and different shapes of scalar non-Gaussianities. 
Such scenarios may even lead to significant non-Gaussian 
corrections to the power spectra along with large non-Gaussian contributions 
to $\ogw$.
Moreover there are efforts to account for the contribution of higher order
non-Gaussianities, such as the trispectrum, to the secondary tensor power 
spectrum. It would be interesting to explore the effects of non-Gaussianities 
with non-trivial scale dependence in such higher order calculations.


\acknowledgments

I thank L.~Sriramkumar for useful discussions at various stages of this
manuscript. I also thank Jens Chluba, V.~Sreenath and Caner Unal for their 
helpful comments.
I thank the Indian Institute of Technology Madras (IIT-M), Chennai, India, 
for financial support through half-time research assistantship.
I acknowledge support from the Science and Engineering Research Board, 
Department of Science and Technology,Government of India, through the Core 
Research GrantCRG/2018/002200.
I acknowledge the use of cluster computing facility at the Department of 
Physics,and the High Performance Computing Environment (HPCE) at IIT-M 
where various numerical computations of this work were carried out.

\appendix

\section{Feynman diagrams for non-Gaussian contributions to $\ogw$}
\label{app:loops}

In order to understand various non-Gaussian contributions to the secondary
GWs, one can construct Feynman diagrams representing these contributions 
(see, for instance, Refs.~\cite{Unal:2018yaa,Atal:2021jyo,Adshead:2021hnm}).
In this appendix, we shall define the elements constituting these diagrams
and present the diagrams corresponding to the contributions we discussed
in Sec.~\ref{sec:ph-2}.

\begin{figure}[b]
\centering
\vskip -10pt
\includegraphics[scale=0.5]{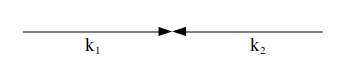}
\includegraphics[scale=0.5]{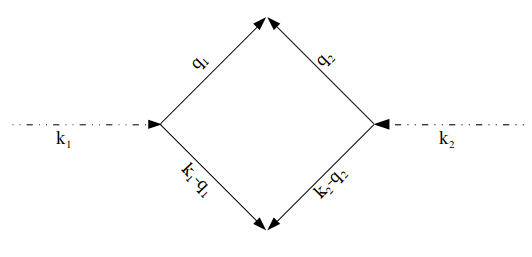}\\
\includegraphics[scale=0.5]{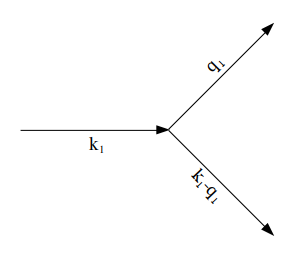}\hspace{2.5cm}
\includegraphics[scale=0.6]{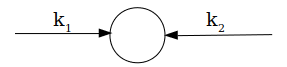}
\caption[Elements of the Feynman diagrams]{The Feynman diagrams representing 
the scalar power spectrum $\ps(k_1)$ (on top left), secondary tensor power 
spectrum $\ph(k_1)$ (on top right) are presented. We also present the diagrams
for the scalar non-Gaussianity parameter $\fnl(k_1,q_1,\vert\vka-\vq_1\vert)$ 
(on bottom left) and the correction to the scalar power $\pc(k_1)$ 
(on bottom right). We use solid lines to represent the scalar mode $\cRk$ 
and dashed-dotted line to represent the secondary tensor mode $h_{\vk}$.}
\label{fig:loop-elem}
\end{figure}
The basic elements that we shall be using for the diagrams are the scalar 
power spectrum $\ps(k)$, secondary tensor power spectrum $\ph(k)$, the 
scalar non-Gaussianity parameter $\fnl(k_1,k_2,k_2)$ and the correction 
to the scalar power spectrum $\pc(k)$.
These diagrams are presented in Fig.~\ref{fig:loop-elem}. Note that the diagram 
representing the secondary tensor power spectrum $\ph(k)$ indicates that it is 
a one loop correction to the primary tensor power spectrum, due to the 
interaction between the tensor and scalar perturbations at the second 
order. The functions $\cI(k,k')$ and $Q^\lambda(k,k')$ arising out of the 
transfer function and polarization tensor, can be accounted at the vertices 
connecting the secondary tensor and scalar modes in the diagram of $\ph(k)$. 
The diagram of $\fnl(k_1,k_2,k_2)$ represents the interaction of scalar 
perturbations $\cRk$ at the cubic order. 
Further, the diagram of $\pc(k)$ indicates that it is a one loop correction 
to the scalar power spectrum $\ps(k)$ due to such cubic order interaction. 
It involves two vertices of $\fnl$ and hence we readily infer that $\pc(k)$ 
shall be proportional to $\fnl^2$.
Using these elements we can construct the diagrams for higher order 
contributions to secondary tensor power spectrum $\ph(k)$ due to scalar 
non-Gaussianity. These shall be higher order loop diagrams arising due to
introduction of the vertex of $\fnl$ in each arm of the loop in the
diagram of $\ph(k)$. 

\begin{figure}
\centering
\includegraphics[scale=0.35]{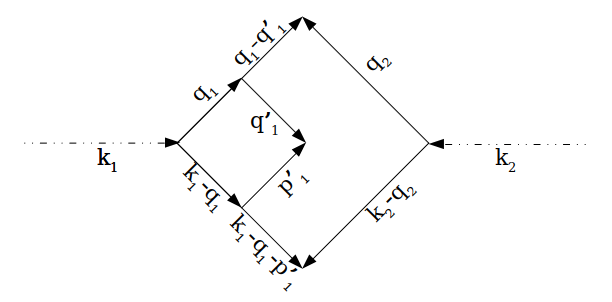}
\includegraphics[scale=0.35]{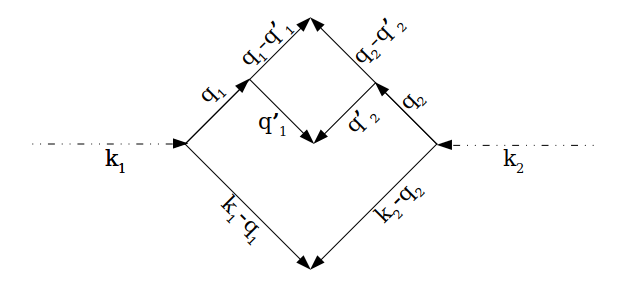}
\includegraphics[scale=0.35]{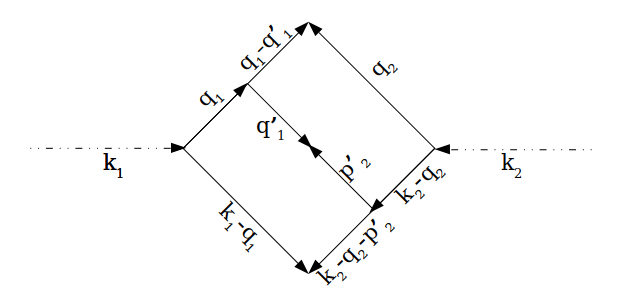}
\caption[Feynman diagrams representing $\ph^{(2-i)}(k)$]{The Feynman 
diagrams representing the non-Gaussian contributions at the level of 
$\fnl^2$ are presented. The term denoted as $\ph^{(2-1)}(k)$ corresponds 
to the C-type diagram (on top left) and the term denoted as $\ph^{(2-2)}(k)$ 
corresponds to the diagram known as the hybrid type (on top right). 
The term denoted as $\ph^{(2-3)}(k)$ corresponds to the Z-type diagram 
(at the bottom).}
\label{fig:ph-2-loops}
\end{figure}
The diagrams representing non-Gaussian contributions to $\ph(k)$ at the level 
of $\fnl^2$, $\ph^{(2-i)}(k)$, are presented in Fig.~\ref{fig:ph-2-loops}.
These diagrams arise due to the introduction of $\fnl$ in two of the four arms 
of the loop in the diagram of $\ph(k)$. They are called as C-type, hybrid and 
Z-type diagrams~\cite{Adshead:2021hnm}.

\begin{figure}
\centering
\includegraphics[scale=0.35]{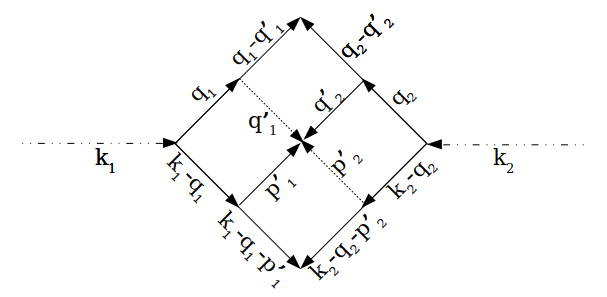}
\includegraphics[scale=0.35]{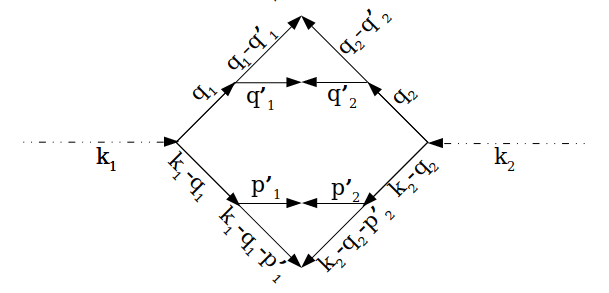}
\includegraphics[scale=0.35]{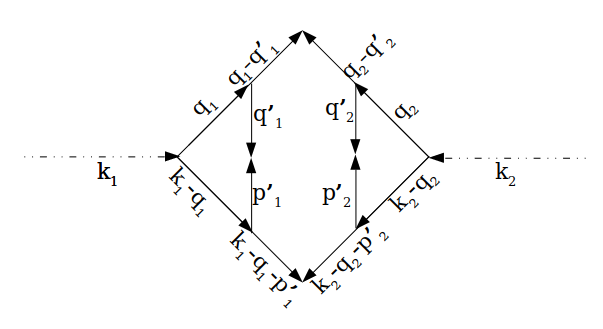}
\caption[Feynman diagrams representing $\ph^{(4-i)}(k)$]{The Feynman 
diagrams representing the non-Gaussian contributions at the level
of $\fnl^4$ are presented. The term denoted as $\ph^{(4-1)}(k)$ 
corresponds to non-planar diagram (on top left). Note that the dotted arrows are 
scalar modes that meet outside the plane of the diagram. The term denoted as 
$\ph^{(4-2)}(k)$ corresponds to the diagram known as reducible term
(on top right). The term denoted as $\ph^{(4-3)}(k)$ 
corresponds to the planar diagram (at the bottom).}\label{fig:ph-4-loops}
\end{figure}
The diagrams representing non-Gaussian contributions to $\ph(k)$ at the level 
of $\fnl^4$ \ie $\ph^{(4-i)}(k)$, are presented in Fig.~\ref{fig:ph-4-loops}.
These diagrams arise when we introduce of $\fnl$ in all the four arms of the 
loop in the diagram of $\ph(k)$. They are called as non-planar, reducible and 
planar diagrams~\cite{Adshead:2021hnm}.

\bibliographystyle{apsrev4-2}
\bibliography{ng_ps_sgw}

\end{document}